\documentclass[%
preprint,
nofootinbib,
aps,
prd,
]{revtex4-1}

\usepackage{graphicx}
\usepackage{psfrag}
\usepackage{float}
\usepackage{subfigure}
\usepackage{array}
\usepackage{graphicx}
\usepackage{amsmath}
\usepackage{amssymb}
\usepackage{mathrsfs}
\usepackage{bm}
\usepackage{slashed}
\usepackage{threeparttable}
\usepackage{multirow}
\usepackage[colorlinks, linkcolor=black, anchorcolor=black, citecolor=black]{hyperref}
\usepackage[amsmath,thmmarks,hyperref]{ntheorem}
\usepackage{soul}
\usepackage{ulem}
\allowdisplaybreaks[4]

\newcommand{\zhou}[1]{{\color{red} #1}}
\newcommand{\ci}{\perp\!\!\!\perp}

\begin{document}


\title{CP violation in non-leptonic $B_c$ decays to excited final states}

\author{Tian Zhou$^1$\footnote{tianzhou@hit.edu.cn}, Tianhong Wang$^1$\footnote{thwang@hit.edu.cn (Corresponding author)}, Hui-Feng Fu$^2$\footnote{huifengfu@jlu.edu.cn}, Zhi-Hui Wang$^3$\footnote{zhwang@nmu.edu.cn}, Lei Huo$^1$\footnote{lhuo@hit.edu.cn}\\ Guo-Li Wang$^{1}$\footnote{gl\_wang@hit.edu.cn}\\}
\address{$^1$School of Physics, Harbin Institute of Technology, Harbin, 150001, China\\
$^2$Center for Theoretical Physics, College of Physics, Jilin University, Changchun 130012, China\\
$^3$School of Electrical and Information Engineering, North Minzu University, Yinchuan 750021, China}


\begin{abstract}

We study the CP violation in two-body nonleptonic decays of $B_c$ meson. We concentrate on the decay channels which contain at least one excited heavy meson in the final states. Specifically, the following channels are considered: $B_c\to c\bar c(2S, 2P)+\bar cq(1S, 1P)$, $B_c\to c\bar c(1S)+\bar cq(2S, 2P)$, $B_c\to c\bar c(1P)+\bar cq(2S)$, $B_c\to c\bar c(1D)+\bar cq(1S, 1P)$, and $B_c\to c\bar c(3S)+\bar cq(1S)$. The improved Bethe-Salpeter method is applied to calculate the hadronic transition matrix element.  Our results show that some decay modes have large branching ratios, which is of the order of $10^{-3}$. The CP violation effect in $B_c \rightarrow \eta_c(1S)+D(2S)$, $B_c \rightarrow \eta_c(1S)+D_0^{*}(2P)$, and $B_c \rightarrow J/\psi+D^{*}(2S)$ are most likely to be found. If the detection precision of the CP asymmetry in such channels can reach the $3\sigma$ level, at least $10^7$ $B_c$ events are needed.   

\end{abstract}

\maketitle


\section{Introduction}

The $B_c$ meson was discovered by CDF more than two decades ago~\cite{Abe:1998fb}. Since then, there have been a lot of studies both theoretical and experimental on this particle. The reason why it's so interesting is that the $B_c$ meson is the lowest bound state which consists of two heavy quarks with different flavors. It cannot decay through strong interaction or electromagnetic interaction, and only the weak decay channels are allowed. This makes the $B_c$ meson an ideal platform to study the properties of heavy quarks and test accurately the Standard Model (SM) predictions.  Experimentally, about $5\times 10^{10}$ $B_c$ mesons per year could be produced at the LHC~\cite{Gouz:2002kk}, which provides us the opportunity to get more information of this particle, such as the CP violation effect in two-body nonleptonic decay modes.

The CP asymmetry in the non-leptonic $B$ decays has been extensively studied by using different methods, such as the QCD factorization approach~\cite{Beneke:1999br,Du:2002up}, the soft-collinear effective theory (SCET)~\cite{Bauer:2004tj,Bauer:2004tj}, and the perturbative QCD (pQCD) method~\cite{Keum:2000ph,Lu:2000em,Li:2003yj}. For $B_c$ meson, the CP violation effects have also attracted some attentions. Based on the pQCD method, Ref.~\cite{Zhang:2009ur, Zou2018, Zhou2012, Xiao:2013lia} studied the direct CP asymmetry parameter of two-body decays of $B_c$ meson with one light final state. In Ref.~\cite{Ivan2003}, a relativistic quark model was applied to explore the decay channels of $B_c^+\to D_s^+\bar D^0$ and $B_c^+\to D_s^+D^0$. In Ref.~\cite{Kar2013}, with a relativistic independent quark model the authors predicted that there were significant CP violation in the $B_c\to D^\ast D^\ast,~D^\ast D_s^\ast$ channels. In Refs.~\cite{Dai:1998hb, Kis2004, Flei2000, Giri2002, Bha2017}, the model-independent method was used to estimate the Cabibbo-Kobayashi-Maskawa (CKM) angle $\gamma$. In Ref.~\cite{Dai:1998hb}, the authors pointed out that to observe the CP violation, about $10^8$ $B_c$ events were needed.

In Refs.~\cite{Fu:2011tn, Chen:2011ut}, the Bethe-Salpeter (BS) method was used to study the CP violation of $B_c$ and $B_s$ mesons. In this formalism, by solving the instantaneous BS equations, we get the wave functions of heavy mesons, which are used to calculate the hadronic transition matrix elements (Ref.~\cite{Liu1997} applied a similar formalism to study the CP violation of $B_c$ with $S$-wave final states). In this paper, we consider the direct CP violation in two-body nonleptonic decays of $B_c$ meson, where a radial excited state is included in the final states. {There are three reasons why we make such study. First, it is still lack of theoretical calculations of these decays. However, as more and more excited $D$ and $D_s$ states are discovered, such as $D_{s1}^*(2710)^{\pm}$, $D(2550)^0$, $D^*(2640)^{\pm}$, and $D_1^*(2680)^0$~\cite{pdg2018}, which can be identified as $D_s^*(2S)$, $D(2S)^0$, $D^*(2S)^{\pm}$, and $D^*(2S)^0$, respectively, those calculations will become necessary. Second, these decay channels can be used to test the validity of the potential models. 
According to Ref.~\cite{Geng:2018qrl}, the relativistic effects are important when the excited states are involved. So the improved BS method~\cite{Zhou:2019stx}, which writes the amplitude in a more covariant form, will be applied. Third, some interesting information can be obtained by studying the CP violation effects. For example, in the decays to the $1^{+(\prime)}$ states, the mixing angle can affect the CP violation severely in some region.}

As the direct CP violation comes from the interference of different diagrams, we will consider the color-favored tree diagram, color-suppressed tree diagram, and the time-like penguin diagram. {The annihilation diagram and the space-like penguin diagram will be helicity suppressed if the masses of the final mesons are small compared with that of the inital meson (this suppression is not very large at this case as the final mesons are also heavy). Besides, the creation of the $c\bar c$ pair also leads to the form factor suppression. So we will not consider these two kinds of diagrams in this work.} The soft strong phase arising from the rescattering effects of final states may also bring considerable contribution. {For example, in Ref.~\cite{Liu:2007qs} the CP asymmetry of the $B_c^+\to D^0\pi^+$ channel is changed about $22\%$ by such effects. To estimate the uncertainties brought by the final state interaction in the channels with two heavy final states is the project of our ongoing work.}

The paper is organized as follows. In Section II, we present the theoretical formalism of the CP asymmetry in nonleptonic decays of $B_c$ meson. In Section III, we use the improved BS method to calculate the hadronic transition matrix elements which are expressed as the overlap integral of the wave functions of heavy mesons. In Section IV,  we give the numerical values of the CP asymmetry and make discussions of the results.

\section{CP violation in non-leptonic decays}

The NLO effective Hamiltonian for $|\Delta B|=1$ nonleptonic decays can be written as~\cite{Buchalla:1995vs}
\begin{equation}\label{eq:effective Hamiltonian}
  \begin{aligned}
    \mathcal{H}_{\mathrm{eff}}(|\Delta B|=1)= \frac{G_{F}}{\sqrt{2}} {\sum_{q=c \atop q^\prime=d,s}} V_{q q^{\prime}} V_{q b}^{*}\left\{C_{1}(\mu) Q_{1}+C_{2}(\mu) Q_{2}+\sum_{k=3}^{10} C_{k}(\mu) Q_{k}\right\}+\mathrm{h.c.},
  \end{aligned}
\end{equation}
where $C_{i}(\mu)(i=1,2, \cdots, 10)$ are the Wilson coefficients and $\mu$ is the renormalization scale. $Q_1$ and $Q_2$ are the tree diagram operators. $Q_{3}$, $Q_{4}$, $Q_{5}$, and $Q_{6}$ are the QCD penguin diagram operators. $Q_{7}$, $Q_{8}$, $Q_{9}$, and $Q_{10}$ are the electroweak penguin diagram operators. Specifically, these operators have the following forms~\cite{Dai:1998hb}
\begin{equation}
  \begin{aligned}
&  Q_{1} =\left(\bar{q}^{\prime}_{\alpha} q_{\beta}\right)_{V-A}\left(\bar{q}_{\beta} b_{\alpha}\right)_{V-A}, \\
&  Q_{2} =\left(\bar{q}^{\prime} q\right)_{V-A}(\bar{q} b)_{V-A}, \\
&  Q_{3(5)} =\left(\bar{q}^{\prime} b\right)_{V-A} \sum_{q^{\prime \prime}}\left(\bar{q}^{\prime \prime} q^{\prime \prime}\right)_{V-A(V+A)}, \\
&  Q_{4(6)} =\left(\bar{q}_{\alpha}^{\prime} b_{\beta}\right)_{V-A} \sum_{q^{\prime \prime}}\left(\bar{q}^{\prime \prime}_{\beta} q_{\alpha}^{\prime \prime}\right)_{V-A(V+A)}, \\
&  Q_{7(9)} =\frac{3}{2}(\overline{q^{\prime}} b)_{V-A} \sum_{q^{\prime \prime}} e_{q^{\prime \prime}}\left(\bar{q}^{\prime \prime} q^{\prime \prime}\right)_{V+A(V-A)}, \\
&  Q_{8(10)} =\frac{3}{2}\left(\overline{q^{\prime}}_{\alpha} b_{\beta}\right)_{V-A} \sum_{q^{\prime \prime}} e_{q^{\prime \prime}}\left(\bar{q}^{\prime \prime}_{\beta} q_{\alpha}^{\prime \prime}\right)_{V+A(V-A)},\\
  \end{aligned}
\end{equation}
where $e_{q^{\prime \prime}}$ is the electric charge of the quark $q^{\prime \prime}$ which can be $u$, $d$, $s$, $c$, or $b$; the subscripts $\alpha$ and $\beta$ are color indices; $(\bar q_{1\alpha}q_{2\beta})_{V \pm A}\equiv \bar q_{1\alpha}\gamma^\mu(1\pm\gamma_5)q_{2\beta}$. As Ref.~\cite{Buras:1991jm} did, we will use the factorization approximation, under which the amplitude can be factorized into the product of hadronic transition matrix element and decay constants. {This is expected to be true when the final mesons are light, so the strong interaction effects can be neglected and the color transparency is valid. If the final mesons are heavy and in the region close to zero recoil, such as the case considered in this work, the factorization approximation may not hold up well, and the final state interactions may give considerable contribution. In Ref.~\cite{Chen:2011ut} both the factorization approximation and the pQCD method are used to study the two-body nonleptonic decays of $B_s$ meson, and the results of two methods are close to each other. This gives us some confidence that the factorization approximation can be applied in these cases, at least as a preliminary estimate.}

\begin{figure}[htb]
 \centering
 \subfigure[~Color-favored tree diagram $\mathcal A_1$]{\includegraphics[scale=0.7]{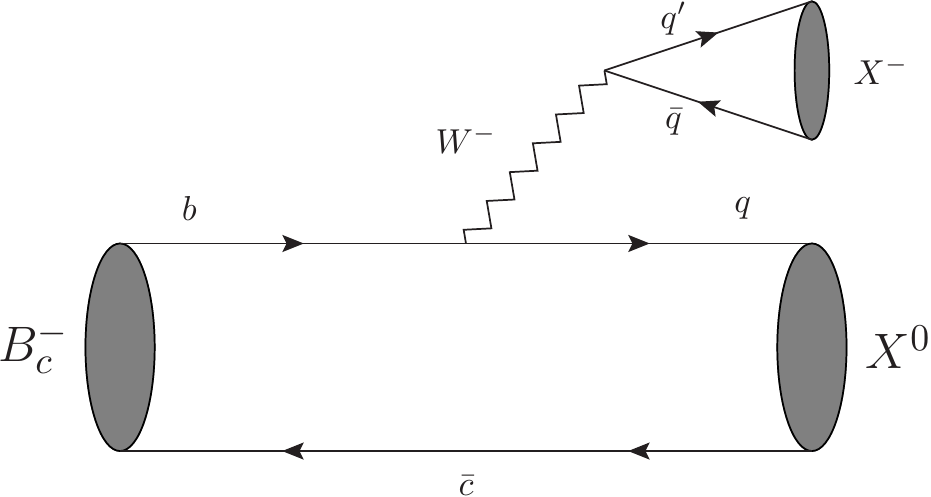}}
 \hspace{1cm}
 \subfigure[~Color-suppressed tree diagram $\mathcal A_2$]{\includegraphics[scale=0.7]{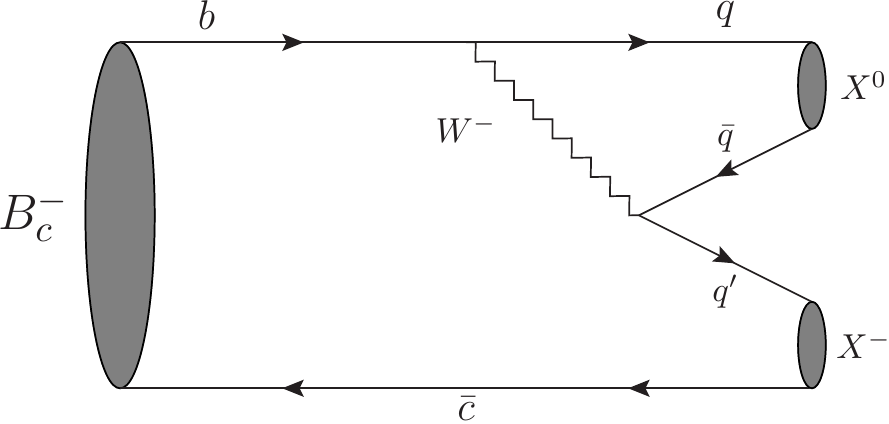}}
 \subfigure[~Time-like penguin diagram $\mathcal A_3$]{\includegraphics[scale=0.7]{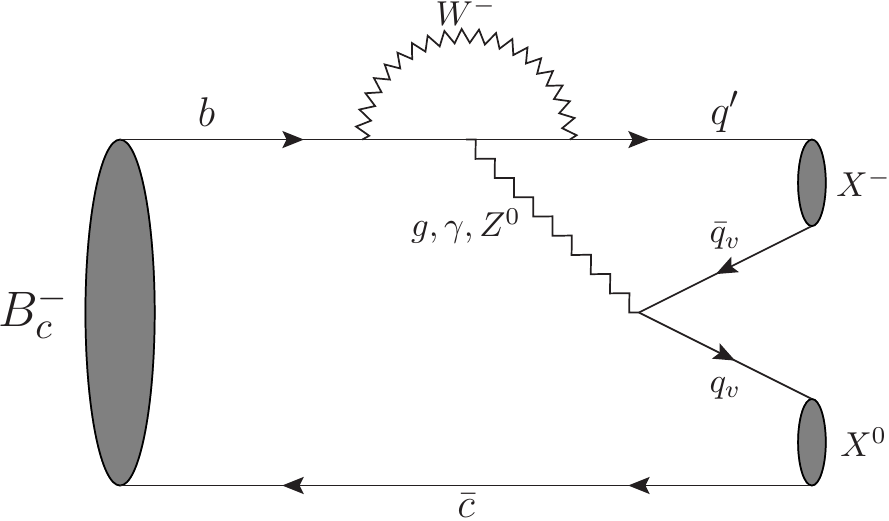}}
 \hspace{1.cm}
 \subfigure[~{Annihilation diagram}]{\includegraphics[scale=0.7]{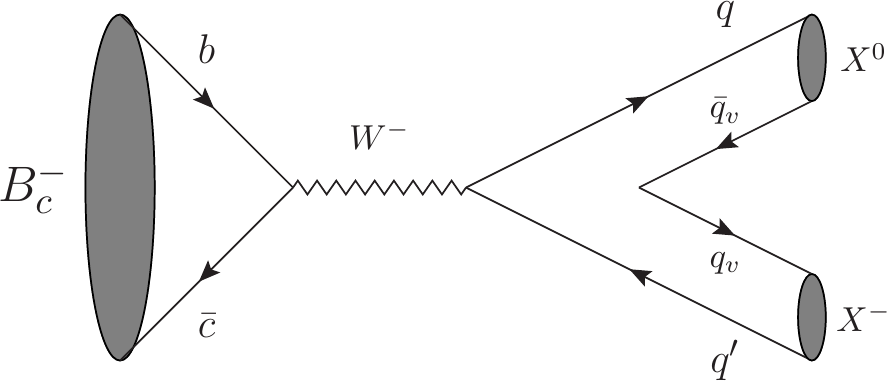}}
  \subfigure[~{Space-like penguin diagram}]{\includegraphics[scale=0.7]{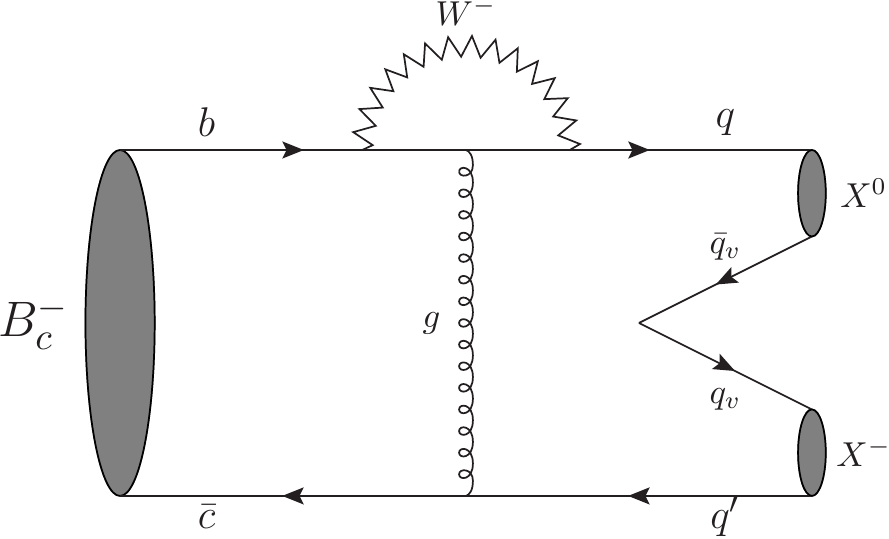}}
\caption{\label{fig:NLO-feynman} The Feynman diagrams for $B_{c}^{-}$ decaying into two mesons $X^{0}$ and $X^{-}$ in the spectator approximation, where $q,~q_{v}=u,~c$  and $q^{\prime},~q_{v}^{\prime}=d,~s$. The subscript $v$ denotes `vacuum'.}
\end{figure}

The amplitudes corresponding to the first three Feynman diagrams in Fig.~\ref{fig:NLO-feynman} are~\cite{Dai:1998hb}
\begin{equation}\label{eq:spectator amplitudes}
  \begin{aligned}
  \mathcal A_1&=a_{1} A, ~~~~~~\mathcal A_2=a_{2} B, \\
  \mathcal A_3&=\left[\left(a_{3}+a_{9}\right)+\xi_{f}\left(a_{5}+a_{7}\right)+\left(1+\xi_{f}\right) b^q G\left(m_q, k^{2}\right)\right] A,
  \end{aligned}
\end{equation}
where A and B are the factorized hadronic matrix elements for $B_{c}^{-} \rightarrow X^{-} X^{0}$,  which have the forms
\begin{equation}
  \begin{aligned}
&  A = \frac{G_F}{\sqrt{2}}\langle X^{-}(q_{1} \bar{q}_{2})\left|\left(\bar{q}_{1} q_{2}\right)_{V-A}\right| 0\rangle\langle X^{0}(q_{3} \bar{c})\left|\left(\bar{q}_{3} b\right)_{V-A}\right| B_{c}^{-}\rangle, \\
&  B = \frac{G_F}{\sqrt{2}}\langle X^{0}(q_{1} \bar{q}_{2})\left|\left(\bar{q}_{1} q_{2}\right)_{V-A}\right| 0\rangle\langle X^{-}(q_{3} \bar{c})\left|\left(\bar{q}_{3} b\right)_{V-A}\right| B_{c}^{-}\rangle.
  \end{aligned}
\end{equation}
The $a_k$ in Eq.~(3) are defined by Wilson coefficients at the renormalization scale $\mu \sim m_{b}$ 
\begin{equation}
a_{2 i-1} \equiv \frac{C_{2 i-1}}{N_{c}}+C_{2 i}, \quad a_{2 i} \equiv \frac{C_{2 i}}{N_{c}}+C_{2 i-1}, \quad(i=1,2,3,4,5)\end{equation}
where $N_c$ is the number of colors. In the actual calculation, one often uses the effective number of colors~\cite{Leitner:2002br}
\begin{equation}
  \frac{1}{\left(N_{c}^{e f f}\right)_{i}}=\frac{1}{3}+\delta_{i}, \text { with } i=1,\cdots,10
\end{equation}
where $\delta_{i}$ stands for the nonfactorizable part. In this paper we will choose $N_c=3$ for all operators to get the results, and we will also discuss the effect of different $N_c^{eff}$.

The parameter $\xi_f$ in the Eq.~\eqref{eq:spectator amplitudes} comes from the Fiertz rearrangement, which transforms the $(V \mp A)(V \pm A)$ currents into the $(S \pm P)(S \mp P)$ currents. For decay channels with different final states, this parameter takes different values~\cite{Chen:2011ut}
\begin{equation}
  \xi_f=\left\{\begin{aligned}
  +\frac{2 M_{B_c}^{2}}{\left(m_{b}-m_{q_v}\right)\left(m_{q^{\prime}}+m_{q_v}\right)} & ~~~~~~~~X_1\left(0^{-}\right) X_2\left(0^{-}\right)~\text {or}~X_1\left(1^{+}\right) X_2\left(0^{-}\right), \\
  -\frac{2 M_{B_c}^{2}}{\left(m_{b}-m_{q_v}\right)\left(m_{q^{\prime}}-m_{q_v}\right)} &~~~~~~~~X_1\left(0^{-}\right) X_2\left(0^{+}\right)~\text {or}~X_1\left(1^{+}\right) X_2\left(0^{+}\right), \\
  -\frac{2 M_{B_c}^{2}}{\left(m_{b}+m_{q_v}\right)\left(m_{q^{\prime}}+m_{q_v}\right)} &~~~~~~~~X_1\left(1^{-}\right) X_2\left(0^{-}\right)~\text {or}~X_1\left(0^{+}\right) X_2\left(0^{-}\right), \\
  +\frac{2 M_{B_c}^{2}}{\left(m_{b}+m_{q_v}\right)\left(m_{q^{\prime}}-m_{q_v}\right)} &~~~~~~~~X_1\left(1^{-}\right) X_2\left(0^{+}\right)~\text {or}~X_1\left(0^{+}\right) X_2\left(0^{+}\right), \\
  \end{aligned}\right.
\end{equation}
where $X_1$ is the final meson which contains the spectator $\bar{c}$ and $X_2$ is the other one. Their $J^P$ numbers are presented in the parentheses. 

The $b^q$ in Eq.~\eqref{eq:spectator amplitudes} have the form
\begin{equation}
  \begin{aligned}
  b^{q}=\left\{\frac{\alpha_{s}}{8 \pi}\left(1-\frac{1}{3 N_{c}}\right) C_{1}+\frac{\alpha_{e}}{3 \pi} \frac{1}{N_{c}}\left(\frac{C_{1}}{N_{c}}+C_{2}\right)\right\}\left[\frac{10}{9}-G\left(m_{q}, k^{2}\right)\right], \\
  \end{aligned}
\end{equation}
where we take $\alpha_{s}(m_Z)=0.1176,~\alpha_{e}(m_{Z})=1/128,~m_{W}=91.1876$ GeV, and $\Lambda_{\mathrm{QCD}}^{(f=5)}=220.9$ MeV. 
The function $G(m_q,k^2)$ is defined as~\cite{Dai:1998hb}
\begin{equation}
G\left(m_q, k^{2}\right)=\frac{3}{2}\left[\frac{10}{9}-F_q\left(k^{2}\right)\right],
\end{equation}
where $F_q(k^2)$ is the penguin loop-integral function of the squared momentum $k^2$ carried by the virtual gluon at the renormalization scale $\mu \sim m_{b}$,
\begin{equation}
F_q\left(k^{2}\right)=-4 \int_{0}^{1} \mathrm{d} x x(1-x) \ln \left[\frac{m_q^2-x(1-x) k^{2}}{m_{b}^{2}}\right].
\end{equation}
According to Ref.~\cite{Gerard:1990ni}, we can rewrite the function $G(m_q, k^2)$ as
\begin{equation}
  \begin{aligned}
  G\left(m_q, k^{2}\right)= \ln \frac{m_q^2}{m_{b}^{2}}-r_q+\left(1+\frac{r_q}{2}\right) \sqrt{1-r_q} \ln \frac{1+\sqrt{1-r_q}}{1-\sqrt{1-r_q}}+i \pi\left(1+\frac{r_q}{2}\right) \sqrt{1-r_q},
  \end{aligned}
\end{equation}
where $r_q=4 m_q^{2} / k^{2}$. As Ref.~\cite{Fu:2011tn} did, we will take $k^2$ to be its average value $\bar k^2$, which is defined as
\begin{equation}
  \frac{\bar{k}^{2}}{m_{b}^{2}}=\frac{1}{2}\left(1+\left(m_{\bar{q}_{v}}^{2}-m_q^{2}\right)\left(1-\frac{m_{\bar q_{v}}^{2}}{m_{b}^{2}}\right) / m_{X^-}^{2}+\left(m_q^{2}+2 m_{\bar{q}_{v}}^{2}-m_{X^-}^{2}\right) / m_{b}^{2}\right).
\end{equation}
In the calculation of the QCD penguin diagram, we will take the current quark masses~\cite{pdg2018}
\begin{equation}\label{eq:current mass}
  \begin{aligned}
 &m_u=0.0022~\text{GeV}, \quad m_d=0.0047~\text{GeV}, \quad m_s=0.095~\text{GeV}, \\
 & m_c=1.275~\text{GeV}, \quad m_b=4.18~\text{GeV}, \quad m_t=173~\text{GeV}.
  \end{aligned}
\end{equation}
Finally the amplitude of two-body nonleptonic decays of $B_c$ meson can be written as~\cite{Dai:1998hb}
\begin{equation}\label{eq:amplitudes}
\mathcal{M}=\frac{G_{F}}{\sqrt{2}}\left\{\left\{\lambda_{c} a_{1}+\sum_{q=u, c} \lambda_q\left[a_{3}+a_{9}+\xi_f\left(a_{5}+a_{7}\right)+(1+\xi_f) b^q\right]\right\} A+\lambda_{c} a_{2} B\right\},
\end{equation}
where $\lambda_q=V_{q b} V_{q d}^{*}$, \zhou{which is accepted for the cause $B_c \to c\bar{c} + cd$ and it should be changed to $\lambda_q=V_{q b} V_{q s}^{*}$ for the channel $B_c \to c\bar{c} + cs$, and the $q$ is the quark inside the QCD penguin}. By combining terms with the same $\lambda_q$, we can write $\mathcal M$ as~\eqref{eq:amplitudes}
\begin{equation}\label{eq:MTT}
\mathcal{M}=V_{c b} V_{c d}^{*} T_{1}+V_{u b} V_{u d}^{*} T_{2}.
\end{equation}

The CP asymmetry of two-body decays of $B_c$ meson is defined as
\begin{equation}
\mathcal{A}_{cp}=\frac{\Gamma\left(B_{c}^{+} \rightarrow \bar{f}\right)-\Gamma\left(B_{c}^{-} \rightarrow f\right)}{\Gamma\left(B_{c}^{+} \rightarrow \bar{f}\right)+\Gamma\left(B_{c}^{-} \rightarrow f\right)}.\end{equation}
By calculating $|\mathcal M(B_{c}^{+} \rightarrow \bar{f})|^2$ and $|\mathcal M(B_{c}^{+} \rightarrow \bar{f})|^2$ and inserting them into Eq.~(16), we get
\begin{equation}\label{eq:CP expression}
\mathcal{A}_{c p}=\frac{-2 {\rm Im}\left(T_{1} T_{2}^{*}\right){\rm Im}\left(\frac{V_{u b} V_{u d}^{*}}{V_{c b} V_{cd}^{*}}\right)}{\left|T_{1}\right|^{2}+\left|\frac{V_{u b} V_{u d}^{*}}{V_{c b} V_{c d}^{*}}\right|^{2}\left|T_{2}\right|^{2}+2 {\rm Re}\left(T_{1} T_{2}^{*}\right){\rm Re}\left(\frac{V_{u b} V_{u d}^{*}}{V_{c b} V_{c d}^{*}}\right)}.
\end{equation}
According to Ref.~\cite{Wolfenstein:1983yz}, the CKM matrix can be parameterized as
\begin{equation}\label{eq:wolfenstein CKM}
  \begin{aligned}
  V =\left(\begin{array}{ccc}
  V_{ud } & V_{us } & V_{u b} \\
  V_{cd} & V_{cs} & V_{c b} \\
  V_{tb} & V_{t s} & V_{t b}
  \end{array}\right) 
  =\left(\begin{array}{ccc}
  1-\frac{\lambda^{2}}{2} & \lambda & A \lambda^{3}(\rho-i \eta) \\
  -\lambda & 1-\frac{\lambda^{2}}{2} & A \lambda^{2} \\
  A \lambda^{3}(1-\rho-i \eta) & -A \lambda^{2} & 1
  \end{array}\right)+O\left(\lambda^{4}\right).
  \end{aligned}
\end{equation}
Inserting Eq.~\eqref{eq:wolfenstein CKM} into Eq.~\eqref{eq:CP expression}, we get
\begin{equation}
  \begin{aligned}
  \mathcal{A}_{c p} &=\frac{-2  {\rm Im}\left(T_{1} T_{2}^{*}\right) \sin \gamma}{\left|T_{1}\right|^{2} / \Delta+\Delta \left|T_{2}\right|^{2}-2  {\rm Re}\left(T_{1} T_{2}^{*}\right) \cos \gamma} \\
  & \equiv D_{1} \frac{\sin \gamma}{1+D_{2} \cos \gamma},
  \end{aligned}
\end{equation}
where we have defined $\gamma \equiv \arg \left(-\frac{V_{u b}^{*} V_{u d}}{V_{c b}^{*} V_{c d}}\right) \simeq \arg \left(\frac{V_{u b}^{*} V_{u s}}{V_{c b}^{*} V_{c s}}\right)$; $\Delta\equiv\left|\frac{V_{u b} V_{u d}^{*}}{V_{c b} V_{c d}^{*}}\right|$. 

\section{The improved BS method}

In this section, we give a brief introduction to the improved BS method, which is used to calculate the hadronic transition matrix element $\left\langle X^{0(-)}\left|J^{V-A}_{\mu}\right| B_{c}\right\rangle$. More details about this method can be found in our previous work~\cite{Zhou:2019stx}. 

According to the Mandelstam formalism, the hadronic matrix can be written as 
\begin{equation}
\left\langle X\left(P_{f}\right)\left|\left(\bar{c} \Gamma^{\mu} b\right)\right| B_{c}(P)\right\rangle=\int \frac{\mathrm{d}^{4} q}{(2 \pi)^{4}} \frac{\mathrm{d}^{4} q_{f}}{(2 \pi)^{4}} \operatorname{Tr}\left[\bar{\chi}_{P_{f}}\left(q_{f}\right) \Gamma^{\mu} \chi_{P}(q) \mathrm{i} S_{2}^{-1}\left(p_{2}\right)\right](2 \pi)^{4} \delta^{(4)}\left(p_{2}-p_{2 f}\right),
\end{equation}
where $\Gamma^{\mu}=\gamma^{\mu}\left(1-\gamma^{5}\right)$; $\chi$ is the BS wave function; $p_i$ ($q$) and $p_{fi}$ ($q_f$) are the (relative) momenta of quark and antiquark within initial and final mesons, respectively; $S_i$ is the quark propagator. By applying the instantaneous approximation, Eq.~(20) can be reduced into the three-dimensional form,
\begin{equation}
\begin{aligned}
\langle D_q (P_{f})|(\overline{c}\Gamma^{\mu} b )|B_q (P)\rangle=-i\int\frac{d\vec q}{(2\pi)^3}{\rm Tr}\left[\frac{\slashed P_f}{M_f}\overline\varphi^{++}(P_f, q_{f\ci})\frac{\slashed P_f}{M_f}L_{r}\Gamma^\mu \varphi^{++}(P, q_\perp)\right],
\end{aligned}
\end{equation}
where
\begin{equation}
  \begin{aligned}
  &\varphi^{++}(P,q_\perp) = \Lambda_1^{+}(p_{1\perp})\frac{\slashed P}{M}\varphi(P, q_\perp)\frac{\slashed P}{M}\Lambda_2^{+}(p_{2\perp}),\\
  &\varphi^{++}(P_f,q_{f\ci}) = \widetilde\Lambda_1^{+}(p_{1f\ci})\frac{\slashed P_f}{M_f}\varphi(P_f, q_{f\ci})\frac{\slashed P_f}{M_f}\widetilde\Lambda_2^{+}(p_{2f\ci})\\
  \end{aligned}
  \end{equation} 
are the positive energy projection of wave functions; the symbol $L_r$ is expressed as
\begin{equation}
L_{r}=\frac{\left(M_{f}-\widetilde\omega_{1f}-\widetilde\omega_{2f}\right)}{\left(P_{f_{P}}-\omega_{1 f}-\omega_2\right)} \Lambda_{1}^{+}\left(p_{1f\perp}\right).
\end{equation}

In the above equations, we have definited the projection operator as follows
\begin{equation}
\begin{aligned}
  &\Lambda^\pm_i(p_{i\perp})=\frac{1}{2\omega_{i}}\left[\frac{\slashed P}{M}\omega_{i} \pm(Jm_{i}+\slashed p_{i\perp})\right],\\
  &\widetilde\Lambda^\pm_i(p_{if\ci})=\frac{1}{2\widetilde\omega_{if}}\left[\frac{\slashed P_f}{M_f}\widetilde\omega_{if} \pm(Jm_{if}+\slashed p_{if\ci})\right],
\end{aligned}
\end{equation}
where $J=1$ and $-1$ for quark and antiquark, respectively; $m_i$ and $m_{if}$ are quark masses. The quark energies $\omega_{i}$, $\omega_{i f}$, and $\widetilde\omega_{i f}$  have the forms
\begin{equation}
  \begin{aligned}
 \omega_{i}=\sqrt{m_{i}^2-q_{\perp}^2},~~~~ \omega_{if}=\sqrt{m_{if}^2-q_{f\perp}^2},~~~\widetilde\omega_{if}=\sqrt{m_{if}^2-q_{f\ci}^2},
  \end{aligned}
\end{equation}
where
\begin{equation}
q_{f\ci}=q_{f\perp}-\frac{q_{f\perp} \cdot P_{f\perp}}{M_{f}^{2}} P_{f}+s_{r}\left(\frac{1}{M} P-\frac{P_{f}\cdot P}{MM_{f}^{2}} P_{f}\right),
\end{equation}
with $s_{r}=\frac{m_2}{m_1+m_2} \frac{P_f\cdot P}{M}-\omega_{2 f}$. The symbols $\perp$ and $\ci$ mean projecting onto the momenta of the initial and final mesons, respectively.

The instantaneous BS wave functions of heavy mesons with different quantum numbers ($0^-$, $0^+$, $1^-$, $1^{+-}$, and $1^{++}$) have the following forms
\begin{equation}
\begin{aligned}
&~~~~~~~~~~\varphi_{0^{-}}(q_{\perp})=M\left[\frac{\slashed P}{M} f_{1}(q_{\perp})+f_{2}(q_{\perp})+\frac{\slashed q_{\perp}}{M} f_{3}(q_{\perp})+\frac{\slashed P\slashed q_\perp}{M^{2}} f_{4}(q_{\perp})\right] \gamma_{5},\\
&~~~~~~~~~~\varphi_{0^{+}}(q_{\perp})=M\left[\frac{\slashed q_{\perp}}{M} g_{1}(q_{\perp})+\frac{\slashed P\slashed q_\perp}{M^{2}} g_{2}(q_{\perp})+g_{3}(q_{\perp})+\frac{\slashed P}{M} g_{4}(q_{\perp})\right], \\
&~~~~~~~~~~\varphi_{1^{-}}(q_{\perp})=(q_{\perp} \cdot \epsilon)\left[h_{1}(q_{\perp})+\frac{\slashed P}{M} h_{2}(q_{\perp})+\frac{\slashed q_{\perp}}{M} h_{3}(q_{\perp})+\frac{\slashed P\slashed q_\perp}{M^{2}} h_{4}(q_{\perp})\right] \\
  &~~~~~~~~~~~~~~~~~~~~~+M \slashed\epsilon\left[h_{5}(q_{\perp})+\frac{\slashed P}{M} h_{6}(q_{\perp})+\frac{\slashed q_{\perp}}{M} h_{7}(q_{\perp})+\frac{\slashed P\slashed q_\perp}{M^{2}} h_{8}(q_{\perp})\right], \\
&~~~~~~~~~~\varphi_{1^{++}}(q_{\perp})=i \varepsilon_{\mu \nu \alpha \beta} \frac{P^{\nu}}{M} q_{\perp}^{\alpha} \epsilon^{\beta}\left[r_{1}(q_{\perp})+\frac{\slashed P}{M} r_{2}(q_{\perp})+\frac{\slashed q_{\perp}}{M} r_{3}(q_{\perp})-\frac{\slashed P\slashed q_\perp}{M^{2}} r_{4}(q_{\perp})\right]\gamma^\mu,\\
& ~~~~~~~~~~\varphi_{1^{+-}}(q_{\perp})=q_{\perp} \cdot \epsilon\left[s_{1}(q_{\perp})+\frac{\slashed P}{M} s_{2}(q_{\perp})+\frac{\slashed q_{\perp}}{M} s_{3}(q_{\perp})+\frac{\slashed P\slashed q_\perp}{M^{2}} s_{4}(q_{\perp})\right] \gamma_{5},
\end{aligned}
\end{equation}
where $f_i$, $g_i$, $h_i$, $r_i$, and $s_i$ are the radial wave functions; $\epsilon^\mu$ is the polarization vector of the vector or axial vector mesons. 
The $1^+(P_1^{1/2})$ and $1^{+\prime}(P_1^{3/2})$ states are the mixing of ${ }^{1} P_{1}$ and ${ }^{3} P_{1}$ states, which can be written as~\cite{Godfrey:2015dva}
\begin{equation}
 \begin{aligned}
 &P_1^{1/2}= {^1P}_{1} \cos \theta_{n P}+{ }^{3} P_{1} \sin \theta_{n P}, \\
 &P_1^{3/2}=-{^1P}_{1} \sin \theta_{n P}+{ }^{3} P_{1} \cos \theta_{n P}.
  \end{aligned}
\end{equation}
According to Ref.~\cite{Godfrey:2015dva}, the mixing angle for different states takes the values:
\begin{equation}\label{eq:mixing angle}
  \begin{aligned}
  \theta_{1P}(c\bar{d})=-25.68^\circ, \quad & \theta_{1P}(c\bar{s})=-37.48^\circ,\\
  \theta_{2P}(c\bar{d})=-29.39^\circ, \quad & \theta_{2P}(c\bar{s})=-30.40^\circ.\\
  \end{aligned}
\end{equation}

\section{Numerical Results and Discussions}

The numerical results of the wave functions are achieved by solving the corresponding instantaneous BS equations.  The Cornell-like potential is adopted, whose detailed forms can be found in Ref.~\cite{Zhou:2019stx}. The values of the related parameters are fixed by fitting the masses of the ground states. The constituent quark masses used in this work, which are different from those of the current quarks when calculating the QCD penguin diagrams, have the values:
\begin{equation}
 \begin{aligned}
 &m_{u}=0.305~\mathrm{GeV}, ~m_{d}=0.311~\mathrm{GeV},~m_{s}=0.500~\mathrm{GeV}, \\
 &m_{c}=1.62~\mathrm{GeV},~m_{b}=4.96~\mathrm{GeV}.
  \end{aligned}
\end{equation}

The PDG values of the CKM matrix elements are~\cite{pdg2018}:
\begin{equation}
 \begin{aligned}
&\left|V_{u d}\right|=0.97425,~\left|V_{u s}\right|=0.2252,~\left|V_{u b}\right|=3.89 \times 10^{-3}, \\
&\left|V_{c d}\right|=0.230,~\left|V_{c b}\right|=0.0406,~\left|V_{c s}\right|=0.97345.
 \end{aligned}
\end{equation}
For the Wilson coefficients, we use the results in Ref.~\cite{Sun:2008wa}:
\begin{equation}
  \begin{aligned}
&C_{1}=-0.1902,\quad C_{2}=1.0849,\quad  C_{3}=0.0148,\quad  C_{4}=-0.0362,\quad C_{5}=0.0088, \\
&C_{6}=-0.0422,\quad  \frac{C_{7}}{\alpha_{e}}=-0.0007,\quad  \frac{C_{8}}{\alpha_{e}}=0.0565,\quad  \frac{C_{9}}{\alpha_{e}}=-1.3039,\quad  \frac{C_{10}}{\alpha_{e}}=0.2700.
  \end{aligned}
\end{equation}
The masses and decay constants of mesons in the ground and excited states are presented in Table~\ref{tab:mass and decay constant in ground}. Here we have used the method in Refs.~\cite{Cvetic:2004qg,Wang:2005qx,Wang:2007av} to calculate the decay constants except those of $D^\pm(1S)$ and $D_s^\pm(1S)$ for which the experimental data are adopted.

\begin{table}[htb]
    \scriptsize
  \begin{ruledtabular}
    \caption{\label{tab:mass and decay constant in ground}  The masses and decay constants of mesons.}
    \begin{tabular}{cccccc}
        particle     & Mass(MeV)~\cite{pdg2018}  & $f_P$(MeV) & particle     & Mass(MeV)~\cite{pdg2018}  & $f_P$(MeV) \\
    \colrule
       $D^\pm(1S)$     & $1869.65 \pm 0.05$  &  $203.2 \pm 5.3 \pm 1.8$~\cite{Ablikim:2013uvu} & $D_{s1}^{\pm}(1P)$  & $2459.5 \pm 0.6$  & 258 \\
       $D_s^\pm(1S)$   & $1968.34 \pm 0.07$  & $241.0 \pm 16.3 \pm 6.5$~\cite{Ablikim:2016duz} & $D^{\prime\pm}_{s1}(1P)$  & $2535.11 \pm 0.06$  &  83.5 \\
       $\eta_c(1S)$  & $2983.9 \pm 0.5$  &  420                                                & $D_{1}^{\pm}(1P)$ & $2423.2 \pm 1.6$  & 283\\
       $D^{*\pm}(1S)$ & $2010.26 \pm 0.05$  & 419                                              & $D^{\prime\pm}_{1}(1P)$  & $2427\pm26\pm25$ &70.2\\
       $D_s^{*\pm}(1S)$& $2112.2 \pm 0.4$ \ & 469                                              & $D_{0}^{*\pm}(1P)$  & $2350.6 \pm 5.9$  &  138 \\
       $J/\psi$     & $3096.900 \pm 0.006$  & 560                                          & $D_{s0}^{*\pm}(1P)$ & $2317.8 \pm 0.5$  &  109   \\
       $B_c(1S)$ & $6274.9 \pm 0.8$             &                                              & $\chi_{c1}(1P)$ & $3510.67 \pm 0.05$  & 239 \\
       \colrule
       particle            & Masses (MeV)  & $f_p$(MeV) & particle            & Masses (MeV)  & $f_p$(MeV) \\
       \colrule
       $D^\pm(2S)$           &  2581~\cite{Godfrey:2015dva}                & 178 & $D_{s1}^\pm(2P)$         &3018~\cite{Godfrey:2015dva}  & 250    \\
       $D_s^\pm(2S)$         &  2673~\cite{Godfrey:2015dva}                & 204 & $D_{s1}^{\prime \pm}(2P)$&3038~\cite{Godfrey:2015dva}  & 58.0   \\
       $\eta_c(2S)$        & $3637.5 \pm 1.1$~\cite{pdg2018}     & 284 & $D_1^\pm(2P)$            &2924~\cite{Godfrey:2015dva}  & 217    \\
       $D^{*\pm}(2S)$        & $2637 \pm 2 \pm 6$~\cite{Abreu:1998vk}      & 325 & $D_1^{\prime \pm}(2P)$   &2961~\cite{Godfrey:2015dva}  & 53.6   \\
       $D_s^{*\pm}(2S)$      & $2732 \pm 4.3 \pm 5.8$~\cite{Aaij:2016utb}  & 372 & $D_0^{*\pm}(2P)$         &2931~\cite{Godfrey:2015dva}  & 79.0   \\
       $\psi(2S)$          & $3686.10 \pm 0.06$~\cite{pdg2018}   & 434 & $D_{s0}^{*\pm}(2P)$      &3005~\cite{Godfrey:2015dva}  & 78.6   \\ 
       $\psi(1D)$          & $3773 \pm 0.4$~\cite{pdg2018}       & 336 & $\chi_{c1}(2P)$          &$3871.69 \pm 0.17$~\cite{pdg2018}  & 229 \\  
       $\psi(3S)$          & $4039 \pm 1$~\cite{pdg2018}         & 375 \\ 
    \end{tabular}
  \end{ruledtabular}
\end{table}

{
By using the numerical results of wave functions and the parameter values mentioned above, we get the form factors of different channels. As an example, we present the form factors $F_{0}^{B_{c} \rightarrow \eta_{c}}$ and $A_{0}^{B_{c} \rightarrow J / \psi}$ at $q^{2}=(P-P_f)^2=0$ in Table \ref{tab: form factor}.  Their expressions can be found in our previous work~\cite{Zhou:2019stx}. For comparison, we also present the results of other models, which are close to ours. 
}

\begin{table}[htbp]
    \scriptsize
\begin{ruledtabular}
  \caption{\label{tab: form factor} { The form factors $F_{0}^{B_{c} \rightarrow \eta_{c}}$ and $A_{0}^{B_{c} \rightarrow J / \psi}$ at $q^{2}=0$.}}
\centering
{
\begin{tabular}{cccccccccc}

 & Ours & pQCD~\cite{pQCD} & SDY~\cite{SDY} & Kiselev~\cite{Kiselev} & IKP~\cite{IKP} & WSL~\cite{WSL} & HZ~\cite{HZ} & DSV~\cite{DSV} & EFG~\cite{EFG}  \\
\hline
$F_{0}^{B_{c} \rightarrow \eta_{c}}$ & 0.577 & 0.72 & 0.66 & 0.66 & 0.79 & 0.61 & 0.87 & 0.58 & 0.47 
\\
$A_{0}^{B_{c} \rightarrow J / \psi}$  & 0.515 & 0.64 & 0.65 &0.60 &0.69 & 0.53 & 0.27 & 0.58 & 0.40 

\end{tabular}}
\end{ruledtabular}
\end{table}

The branching ratios ($Br$) and CP asymmetry ($\mathcal A_{cp}$) of different decay channels are presented in Table~\ref{tab:Bc to 2S 1S}, Table~\ref{tab:Bc to 1S 2S}, and Table~\ref{tab:Bc to 1D 1S}. {The uncertainties come from varying the parameter values simultaneously by $\pm 5\%$ when solving the instantaneous BS equations.} The values of $D_{1, 2}$ defined in Eq.~(19) are also given. Following Refs.~\cite{Fu:2011tn, Dai:1998hb}, we also give an estimation of how many $B_c$ events are needed to observe the CP violation effect. If it is observed at three standard deviation ($3\sigma$) level, the $B_c^{\pm}$ events needed are $\epsilon_{f} N \sim \frac{9}{B r \mathcal{A}_{c p}^{2}}$, where $\epsilon_{f}$ is the detecting efficiency of the final state~\cite{Fu:2011tn}. For the cases when the radial excited state is the charmonium (Table~\ref{tab:Bc to 2S 1S} and~\ref{tab:Bc to 1D 1S}), one can see that the $\psi(2S) D_{s}^{*}$ channel has the largest branching ratio but a small $\mathcal A_{cp}$; the $\psi(1D) D_{0}^{*}$ channel has the largest $\mathcal A_{cp}$ but a small branching ratio; the $\epsilon_f N$ for $\eta_c(2S) D(1S)$ and $\psi(2S) D^*(1S)$ channels are of the order of $10^8$, which is possible to be observed by the current experiments. For the cases when the radial excited state is the heavy-light meson (Table~\ref{tab:Bc to 1S 2S}), there are five channels whose $\epsilon_f N$ is of the order of $10^7\sim10^8$: $B_c \rightarrow \eta_c(1S) D(2S)$, $B_c \rightarrow \eta_c(1S) D_0^*(2P)$, $B_c \rightarrow \eta_c(1S) D^*(2S)$, $B_c \rightarrow \eta_c(1S) D_s(2S)$, $B_c \rightarrow J/\psi D^*(2S)$. One also notices that the decay channels $J / \psi D_{s}^{*}(2S)$ and $\eta_{c} D^{}(2S)$ have the largest branching ratio and $\mathcal A_{cp}$, respectively.

\begin{figure}[htb]
  \centering
  \includegraphics[scale=0.33]{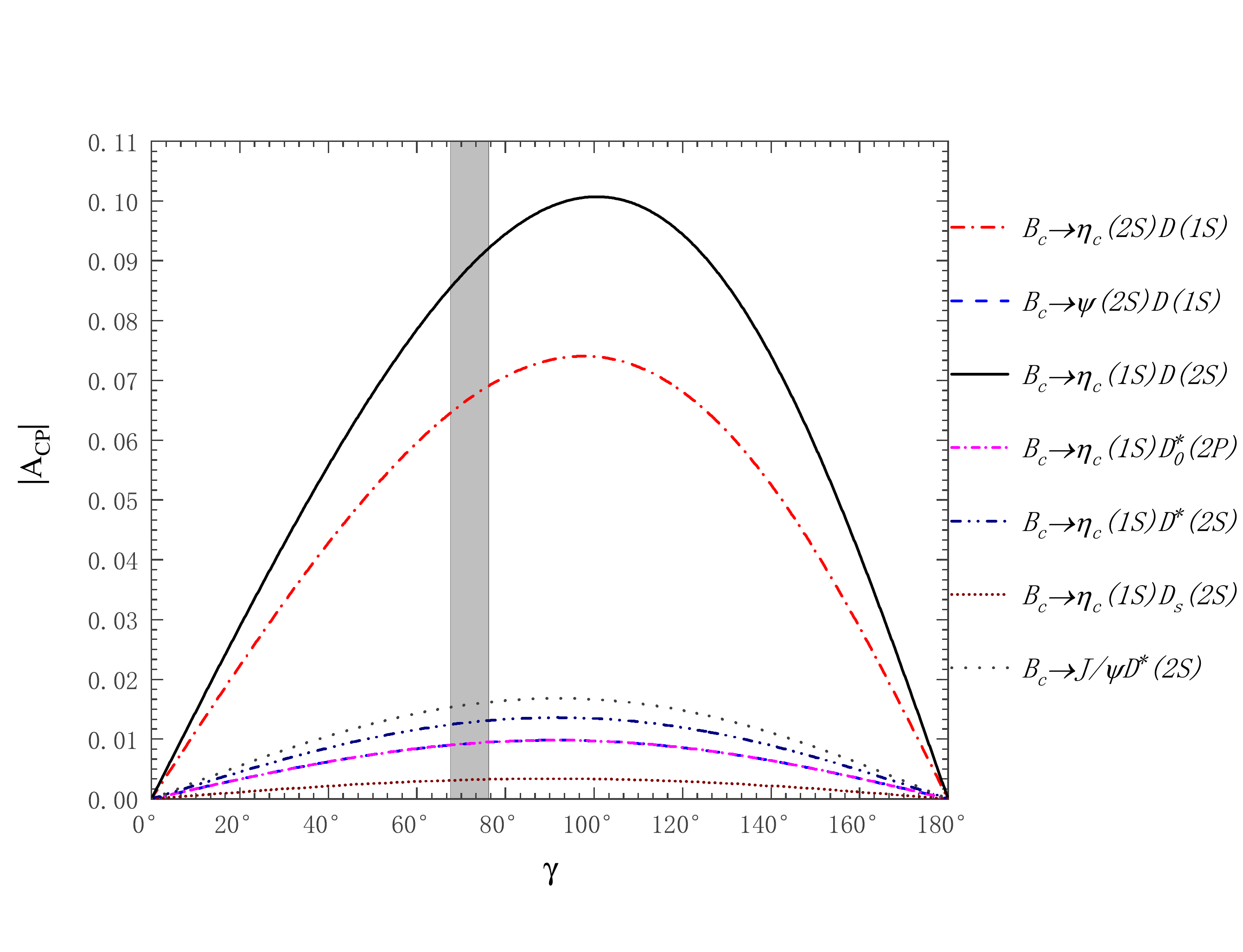}
  \caption{\label{fig: CP-gamma} The dependence of $|\mathcal{A}_{C P}|$ on the weak phase $\gamma$. The gray band is the constrained range of $\gamma$ measured in $B$ decay, which is about $67.6^{\circ} \sim 76.2^{\circ}$~\cite{pdg2018}.}
\end{figure}

The CP asymmetry is related to the weak CP phase $\gamma$ by Eq.~(19).  Experimentally, $\gamma$ is constrained by the nonleptonic decays of $B$ meson, the latest results of which in PDG2020 is {$(72.1^{+4.1}_{-4.5})^\circ$}~\cite{pdg2018}. In Fig.~\ref{fig: CP-gamma} we draw $|\mathcal A_{cp}|$ as a function of $\gamma$. The eight decay channels, whose $|\mathcal A_{cp}|$ are of the order of $10^7\sim10^8$, have been plotted. The maxima of the curves are achieved when $\gamma$ takes about $100^\circ$. The gray band indicates the experimental upper and lower limits of $\gamma$. So when we calculate $\mathcal A_{cp}$ of these channels, $\gamma$ will bring errors to the results. We can see that for the $B_c \rightarrow \eta_c(2S)D(1S)$ and $B_c \rightarrow \psi(2S)D(1S)$ channels, $\mathcal A_{cp}$ changes about {$7\%$}, while for the other channels, the changes of $\mathcal A_{cp}$ are very small.

Next we study how the results are affected by the effective color number $N_c^{eff}$ and the squared momentum $k^2$ carried by the virtual gluon. According to Refs.~\cite{Leitner:2002br,Gerard:1988jj}, $k^2$ varies in the ranges $[m_b^2/4,~m_b^2/2]$ or $[0,~m_b^2]$. So here we will let $k^2$ take four different values: $0.35m_b^2$, $0.5m_b^2$, $0.65m_b^2$, and $0.8m_b^2$. Besides, the $N_c^{eff}$ will vary from 2 to 10, which accounts for the errors brought by the non-factorization effect. The results are shown in Figs.~\ref{fig: bcnc1sD2s} $\sim$ \ref{fig: bcu2ssD1sstar}. Here we only consider the decay channels with $\epsilon_f N\sim10^7$ or $10^8$. One can see that as $N_c^{eff}$ changes, $\mathcal A_{cp}$ and $Br$ can change at most several times. The changes of $k^2$ have even little effect on these quantities. For example, the branching ratio of $\eta_c(2S)D_0^*(1S)$ channel (see Fig.~\ref{fig: bcnc2sD1sstar}) changes less than $50\%$ when $k^2$ changes from $0.35m_b^2$ to $0.80m_b^2$.

At last, we study how the $P$-wave mixing angle affects the CP asymmetry.  The results are shown in Fig.~10 and Fig.~11. One can see that except $B_c\to J/\psi D_{s1}^{(\prime)}(2P)$, all the other channels with a $1^{+(\prime)}$ final state have a critical angle around which $\mathcal A_{cp}$ changes severely.

In conclusion, we have calculated the CP violation in two-body nonleptonic decays of $B_c$, where excited states are included in the final states. Some decay modes have large branching ratios, which is of the order of $10^{-3}$. We studied in detail seven decay channels whose CP asymmetry could be detected by the current experiments. Among these channels, $B_c \rightarrow \eta_c(1S)D(2S)$, $B_c \rightarrow \eta_c(1S)D_0^{*}(2P)$, and $B_c \rightarrow J/\psi D^{*}(2S)$ are the most promising ones, for which, about $10^7$ $B_c$ events are needed, if the CP violation is observed at $3\sigma$ level.  

\begin{table}[htb]
  \scriptsize
  \begin{ruledtabular}
    \caption{\label{tab:Bc to 2S 1S}  The results for $B_{c} \rightarrow c\bar c(2S, 2P)+\bar cq(1S,1P)$. }
    \begin{tabular}{ccccccc}
      Channel     &$D_1$   &  $D_2$  & $Br(B_c^+)$ & $Br(B_c^-)$ &$\mathcal{A}_{c p}$\%& $\epsilon_fN$   \\
      \colrule
      $\eta_{c}(2S) D^{}$ & 0.0734 & 0.133 & $3.64^{+0.54}_{-0.36} \times 10^{-6}$  & $3.19^{+0.54}_{-0.36} \times 10^{-6}$ & $6.53^{+0.09}_{-0.08}$ & $(5.19 \sim 7.08) \times 10^{8}$
      \\
      $\eta_{c}(2S) D_{0}^{*}$ & 0.0349 & 0.0723 & $6.22^{+0.21}_{-0.27} \times 10^{-6}$  & $5.84^{+0.21}_{-0.27} \times 10^{-6}$ & $3.17^{+0.11}_{-0.15}$ & $(1.33 \sim 1.71) \times 10^{9}$
      \\
      $\eta_{c}(2S) D^{*}$ & 0.0123 & 0.0235 & $2.92^{+0.18}_{-0.14} \times 10^{-5}$  & $2.85^{+0.18}_{-0.14} \times 10^{-5}$ & $1.13^{+0.07}_{-0.06}$ & $(2.04\sim 2.85) \times 10^{9}$
      \\
      $\eta_{c}(2S) D_1^{}$ & {0.00148} & {0.00299} & {$8.58^{+0.01}_{-0.02}\times10^{-9}$} & {$8.56^{+0.01}_{-0.02} \times 10^{-9}$} & {$0.138^{+0.001}_{-0.001}$} & {$(5.42 \sim 5.60) \times 10^{12}$}
      \\
      $\eta_{c}(2S) D_1^{\prime}$ & {0.0213} & {0.0412} & {$9.39^{+0.05}_{-0.09} \times 10^{-7}$} & {$9.04^{+0.05}_{-0.09} \times 10^{-7}$} & {$1.96^{+0.02}_{-0.04}$} & {$(2.39 \sim 2.57) \times 10^{10}$}
      \\
      $\eta_{c}(2S) D_{s}^{}$ & -0.00156 & -0.00316 & {$4.46^{+0.01}_{-0.02} \times 10^{-4}$} & {$4.47^{+0.01}_{-0.02} \times 10^{-4}$}  & $-0.145^{+0.003}_{-0.005}$ & $(9.17 \sim 10.33) \times 10^{9}$
      \\
      $\eta_{c}(2S) D_{s 0}^{*}$ & -0.0171 & -0.0228 & $2.49^{+0.08}_{-0.09} \times 10^{-6}$ & $2.57^{+0.08}_{-0.09} \times 10^{-6}$  & $-1.61^{+0.06}_{-0.05}$ & $(1.33 \sim 1.43) \times 10^{10}$
      \\
      $\eta_{c}(2S) D_{s}^{*}$ & -0.000634 & -0.00122 & $3.56^{+0.02}_{-0.02} \times 10^{-4}$ & $3.56^{+0.02}_{-0.02} \times 10^{-4}$ & $-0.0591^{+0.0003}_{-0.0003}$ & $(7.17 \sim 7.32) \times 10^{10}$
      \\
      $\eta_{c}(2S) D_{s 1}$ & {-0.000186} & {-0.000375} & $1.76^{+0.07}_{-0.06} \times 10^{-5}$ & $1.76^{+0.07}_{-0.06} \times 10^{-5}$ & {$-0.0173^{+0.0012}_{-0.0011}$} & $(1.44 \sim 2.02) \times 10^{13}$
      \\
      $\eta_{c}(2S) D_{s 1}^{\prime}$ & -0.00135 & -0.00261 & $3.46^{+0.18}_{-0.07} \times 10^{-6} $ & $3.47^{+0.18}_{-0.07} \times 10^{-6}$ & $-0.126^{+0.004}_{-0.005}$ & $(1.55 \sim 1.74) \times 10^{12}$
      \\
      \colrule
      $\psi(2S) D^{}$ & 0.00431 & 0.000567 & $1.56^{+0.01}_{-0.05} \times 10^{-6}$ & $1.55^{+0.01}_{-0.05} \times 10^{-6}$ & $0.402^{+0.003}_{-0.001}$ & $(3.31 \sim 4.07) \times 10^{11}$
      \\
      $\psi(2S) D_{0}^{*}$ & -0.00199 & -0.00712 & $2.52^{+0.08}_{-0.09} \times 10^{-6}$ & $2.53^{+0.08}_{-0.09} \times 10^{-6}$ & $-0.186^{+0.024}_{-0.034}$ & $(0.82 \sim 1.12) \times 10^{12}$
      \\
      $\psi(2S) D^{*}$ & 0.00981 & 0.0189 & $1.29^{+0.12}_{-0.23} \times 10^{-4}$ & $1.26^{+0.12}_{-0.23} \times 10^{-4}$ & $0.908^{+0.071}_{-0.058}$ & $(6.88 \sim 10.44) \times 10^{8}$
      \\
      $\psi(2S) D_1^{}$ & {0.00177} & {0.00357} & {$3.70^{+0.02}_{-0.01} \times 10^{-6}$} & {$3.69^{+0.02}_{-0.01} \times 10^{-6}$} & {$0.164^{+0.001}_{-0.001}$} & {$(0.98 \sim 1.02) \times 10^{10}$}
      \\
      $\psi(2S) D_1^{\prime}$ &{0.0204} & {0.0394} & {$6.42^{+0.08}_{-0.04} \times 10^{-6}$} & {$6.18^{+0.08}_{-0.04} \times 10^{-6}$} & {$1.88^{+0.01}_{-0.02}$} & {$(3.83 \sim 4.03) \times 10^{9}$}
      \\
      $\psi(2S) D_{s}^{}$ & -0.0000893 & -0.0000238 & {$2.69^{+0.03}_{-0.03} \times 10^{-4}$} & {$2.69^{+0.03}_{-0.03} \times 10^{-4}$} & $-0.00832^{+0.00020}_{-0.00020}$ & $(3.13 \sim 8.37) \times 10^{12}$
      \\
      $\psi(2S) D_{s 0}^{*}$ & -0.00252  & -0.00682 & $2.82^{+0.08}_{-0.06} \times 10^{-7}$ & $2.83^{+0.08}_{-0.06} \times 10^{-7}$ & $-0.236^{+0.005}_{-0.004}$ & $(5.65 \sim 5.80) \times 10^{12}$
      \\
      $\psi(2S) D_{s}^{*}$ & -0.000515 & -0.00100 & {$2.72^{+0.05}_{-0.07} \times 10^{-3} $} & {$2.73^{+0.05}_{-0.07} \times 10^{-3}$} & $-0.0481^{+0.0007}_{-0.0005}$ & $(1.21 \sim 1.60) \times 10^{10}$
      \\
      $\psi(2S) D_{s 1}$ & {-0.000314} & {-0.000631} & {$1.15^{+0.13}_{-0.10} \times 10^{-5}$} & {$1.15^{+0.13}_{-0.10} \times 10^{-5}$} & {$-0.0293^{+0.0031}_{-0.0029}$} & {$(6.69 \sim 12.3) \times 10^{11}$}
      \\
      $\psi(2S) D_{s 1}^{\prime}$ & {-0.00133} & {-0.00256} & {$3.43^{+0.93}_{-0.60}\times 10^{-5}$} & {$3.44^{+0.93}_{-0.60} \times 10^{-5}$} & {$-0.124^{+0.031}_{-0.031}$} & {$(1.41 \sim 1.96) \times 10^{11}$}
      \\
      \colrule
      $\chi_{c1}(2P) D^{}$ & -0.0356 & -0.0878 & $8.23^{+4.23}_{-2.38} \times 10^{-8}$ & $8.81^{+4.23}_{-2.38} \times 10^{-8}$ & $-3.43^{+1.32}_{-1.13}$ & $(3.13 \sim 27.7) \times 10^{10}$
      \\
      $\chi_{c1}(2P) D_{0}^{*}$ & -0.0164 & -0.0376 & $4.27^{+0.29}_{-0.80} \times 10^{-8}$ & $4.19^{+0.29}_{-0.80} \times 10^{-8}$ & $1.00^{+0.22}_{-0.21}$ & $(0.42 \sim 13.4) \times 10^{12}$
      \\
      $\chi_{c1}(2P) D^{*}$ & 0.0246 & 0.0428 & $2.49^{+0.12}_{-0.08} \times 10^{-5}$ & $2.45^{+0.12}_{-0.08} \times 10^{-5}$ & $0.781^{+0.042}_{-0.030}$ & $(5.16 \sim 6.69)  \times 10^{9}$
      \\
      $\chi_{c1}(2P) D_{s}^{}$ & -0.000734 & -0.00154 &$1.68^{+0.11}_{-0.06} \times 10^{-5}$ & $1.69^{+0.11}_{-0.06} \times 10^{-5}$ & $-0.0685^{+0.0122}_{-0.0189}$ & $(0.72 \sim 1.58) \times 10^{12}$
      \\
      $\chi_{c1}(2P) D_{s 0}^{*}$ & 0.000632 & 0.00149 & $8.06^{+1.81}_{-0.38} \times 10^{-7}$ & $8.05^{+1.81}_{-0.38} \times 10^{-7}$ & $0.0589^{+0.0114}_{-0.0031}$ & $(1.84 \sim 3.76) \times 10^{13}$
      \\
      $\chi_{c1}(2P) D_{s}^{*}$ & -0.000381 & -0.000744 & $3.76^{+0.03}_{-0.05} \times 10^{-4} $ & $3.77^{+0.03}_{-0.05} \times 10^{-4}$ & $-0.0355^{+0.0021}_{-0.0011}$ & $(1.67 \sim 2.05) \times 10^{11}$
      \\
    \end{tabular}
  \end{ruledtabular}
\end{table}

\begin{table}[htb]
    \scriptsize
  \begin{ruledtabular}
    \caption{\label{tab:Bc to 1S 2S}  The results for $B_{c} \rightarrow c\bar c(1S)+\bar cq(2S,2P)$ and $B_{c} \rightarrow c\bar c(1P)+\bar cq(2S)$.}
    \begin{tabular}{ccccccc}
      Channel     &$D_1$   &  $D_2$  & $Br(B_c^+)$ &$Br(B_c^-)$  &$\mathcal{A}_{c p}$\%& $\epsilon_fN$   \\
      \colrule
      $\eta_{c} D^{}(2S)$ & 0.0990 & 0.184 & $4.01^{+0.06}_{-0.06} \times 10^{-5}$ & $3.37^{+0.06}_{-0.06} \times 10^{-5}$ & $8.65^{+0.13}_{-0.14}$ & $(3.11 \sim  3.42) \times 10^{7}$
      \\
      $\eta_{c} D_{0}^{*}(2P)$ & 0.0982 & 0.190 & $3.44^{+0.06}_{-0.04} \times 10^{-5}$ & $2.90^{+0.06}_{-0.04} \times 10^{-5}$ & $8.56^{+0.15}_{-0.10}$ & $(3.59 \sim 3.93) \times 10^{7}$
      \\
      $\eta_{c} D^{*}(2S)$ & 0.0136 & 0.0270 & $1.83^{+0.11}_{-0.34} \times 10^{-4}$ & $1.78^{+0.11}_{-0.34} \times 10^{-4}$ & $1.25^{+0.08}_{-0.32}$ & $(2.65 \sim 7.10) \times 10^{8}$
      \\
      $\eta_{c} D_1^{}(2P)$ & {0.00636} & {0.0132} & {$1.38^{+0.56}_{-0.09} \times 10^{-5}$} & {$1.36^{+0.56}_{-0.09} \times 10^{-5}$} & {$0.590^{+0.211}_{-0.051}$} & {$(0.72 \sim 2.42) \times 10^{10}$}
      \\
      $\eta_{c} D_1^{\prime}(2P)$ & {0.0182} & {0.0361} & {$3.05^{+0.06}_{-0.04} \times 10^{-5}$} & {$2.95^{+0.06}_{-0.04} \times 10^{-5}$} & {$1.67^{+0.02}_{-0.03}$} & {$(1.02 \sim 1.13) \times 10^{9}$}
      \\
      $\eta_{c} D_{s}^{}(2S)$ & -0.00337 & -0.00677 & $2.65^{+0.06}_{-0.06} \times 10^{-3}$ & $2.66^{+0.06}_{-0.06} \times 10^{-3}$  & $-0.315^{+0.004}_{-0.004}$ & $(3.40 \sim 3.43) \times 10^{8}$
      \\
      $\eta_{c} D_{s 0}^{*}(2P)$ & -0.00841 & -0.0147 & $9.23^{+0.26}_{-0.36} \times 10^{-5}$ & $9.38^{+0.26}_{-0.36} \times 10^{-5}$  & $-0.788^{+0.032}_{-0.022}$ & $(1.53 \sim 1.64) \times 10^{9}$
      \\
      $\eta_{c} D_{s}^{*}(2S)$ & -0.000738 & -0.00147 & $2.31^{+0.25}_{-0.26} \times 10^{-3}$ & $2.31^{+0.25}_{-0.26} \times 10^{-3}$ & $-0.0688^{+0.0027}_{-0.0021}$ & $(6.87 \sim 10.4) \times 10^{9}$
      \\
      $\eta_{c} D_{s 1}(2P)$ & -0.000704 & -0.00143 & $6.24^{+0.05}_{-0.05} \times 10^{-5}$ & $6.25^{+0.05}_{-0.05} \times 10^{-5}$  & $-0.0656^{+0.0042}_{-0.0052}$ & $(2.76 \sim 4.15) \times 10^{11}$
      \\
      $\eta_{c} D_{s 1}^{\prime}(2P)$ & -0.000861 & -0.00174 & $3.61^{+0.52}_{-0.41} \times 10^{-4}$ & $3.62^{+0.52}_{-0.41} \times 10^{-4}$ & $-0.0803^{+0.0002}_{-0.0002}$ & $(3.82 \sim 3.89) \times 10^{10}$
      \\
      \colrule
      $J / \psi D^{}(2S)$ & -0.0105 & -0.0324 & $2.43^{+0.05}_{-0.07} \times 10^{-5}$ & $2.48^{+0.05}_{-0.07} \times 10^{-5}$ & $-0.989^{+0.003}_{-0.002}$ & $(3.69 \sim 3.84) \times 10^{9}$
      \\
      $J / \psi D_{0}^{*}(2P)$ & -0.0163 & -0.0475 & $4.56^{+0.19}_{-0.26} \times 10^{-6}$ & $4.70^{+0.19}_{-0.26} \times 10^{-6}$ & $-1.55^{+0.09}_{-0.06}$ & $(7.95 \sim 8.76) \times 10^{9}$
      \\
      $J / \psi D^{*}(2S)$ & 0.0168 & 0.0332 & $4.93^{+1.37}_{-0.58} \times 10^{-4}$ & $4.78^{+1.37}_{-0.58} \times 10^{-4}$ & $1.55^{+0.34}_{-0.21}$ & $(0.40 \sim 1.17) \times 10^{7}$
      \\
      $J / \psi D_1^{}(2P)$ & {0.00912} & {0.0185} & {$5.45^{+0.27}_{-0.18} \times 10^{-5}$} & {$5.36^{+0.27}_{-0.18} \times 10^{-5}$} & {$0.845^{+0.028}_{-0.039}$} & {$(2.08 \sim 2.65) \times 10^{9}$}
      \\
      $J / \psi D_1^{\prime}(2P)$ & {0.0164} & {0.0326} & {$3.07^{+0.17}_{-0.27} \times 10^{-4}$} & {$2.98^{+0.17}_{-0.27} \times 10^{-4}$} & {$1.51^{+0.15}_{-0.08}$} & {$(9.01 \sim 10.19) \times 10^{9}$}
      \\
      $J / \psi D_{s}^{}(2S)$ & 0.000347 & 0.00109 &$2.54^{+0.02}_{-0.01} \times 10^{-3}$ & $2.53^{+0.02}_{-0.01} \times 10^{-3}$ & $0.0347^{+0.0002}_{-0.0001}$ & $(2.89 \sim 2.97) \times 10^{10}$
      \\
      $J / \psi D_{s 0}^{*}(2P)$ & 0.00126 & 0.00372 & $2.62^{+0.04}_{-0.03} \times 10^{-5}$ & $2.61^{+0.04}_{-0.03} \times 10^{-5}$ & $0.118^{+0.003}_{-0.002}$ & $(2.31 \sim 2.58) \times 10^{11}$
      \\
      $J / \psi D_{s}^{*}(2S)$ & -0.000738 & -0.00147 & {$1.75^{+0.13}_{-0.02} \times 10^{-2} $} & $1.75^{+0.13}_{-0.02} \times 10^{-2}$ & $-0.0688^{+0.0027}_{-0.0021}$ & $(0.94 \sim 1.17) \times 10^{9}$
      \\
      $J / \psi D_{s 1}(2P)$ & -0.000309 & -0.000629 & $1.90^{+0.03}_{-0.02} \times 10^{-3}$ & $1.90^{+0.03}_{-0.02} \times 10^{-3}$  & $-0.0288^{+0.0022}_{-0.0042}$ & $(4.61 \sim 7.90) \times 10^{10}$
      \\
      $J/ \psi D_{s 1}^{\prime}(2P)$ & -0.000805 & -0.00163 & $6.32^{+0.09}_{-0.18} \times 10^{-3}$ & $6.33^{+0.09}_{-0.18} \times 10^{-3}$ & $-0.0751^{+0.0023}_{-0.0011}$ & $(2.34 \sim 2.67) \times 10^{9}$
      \\
      \colrule
      $\chi_{c1} D^{}(2S)$ & 0.337 & 0.408 & $11.50^{+0.45}_{-0.38} \times 10^{-9}$ & $6.55^{+0.45}_{-0.38} \times 10^{-9}$ & {$27.4^{+1.3}_{-1.2}$} & $(0.57 \sim 4.44) \times 10^{10}$
      \\
      $\chi_{c1} D^{*}(2S)$ & 0.0127 & 0.0253 & $2.43^{+0.43}_{-0.22} \times 10^{-5}$ & $2.37^{+0.43}_{-0.23} \times 10^{-5}$ & $1.17^{+0.19}_{-0.13}$ & $(1.71 \sim 3.83) \times 10^{9}$
      \\
      $\chi_{c1} D_{s}^{}(2S)$ & -0.00225 & -0.00473 &$2.53^{+0.71}_{-0.92} \times 10^{-6}$ & $2.54^{+0.71}_{-0.92} \times 10^{-6}$ & $-0.210^{+0.112}_{-0.043}$ & $(2.67 \sim 19.98) \times 10^{11}$
      \\
      $\chi_{c1} D_{s}^{*}(2S)$ & -0.000986 & -0.00194 & $1.29^{+0.45}_{-0.28} \times 10^{-4} $ & $1.29^{+0.45}_{-0.28} \times 10^{-4}$ & $-0.0920^{+0.0221}_{-0.0201}$ & $(7.08 \sim 10.58) \times 10^{10}$
      \\
    \end{tabular}
  \end{ruledtabular}
\end{table}

\begin{table}[htb]
    \scriptsize
  \begin{ruledtabular}
    \caption{\label{tab:Bc to 1D 1S}  The results for $B_{c} \rightarrow c\bar c(1D)+\bar cq(1S,1P)$ and $B_{c} \rightarrow c\bar c(3S)+\bar cq(1S)$.}
    \begin{tabular}{ccccccc}
      Channel     &$D_1$   &  $D_2$  & $Br(B_c^+)$ & $Br(B_c^-)$ &$\mathcal{A}_{c p}$\%& $\epsilon_fN$   \\
      \colrule
      $\psi(1D) D^{}$ & 0.00160 & 0.000216 & $8.25^{+0.14}_{-0.32} \times 10^{-6}$ & $8.23^{+0.14}_{-0.32} \times 10^{-6}$ & $0.150^{+0.002}_{-0.005}$ & $(4.65 \sim 5.40) \times 10^{11}$
       \\
       $\psi(1D) D_{0}^{*}$ & 0.0890 & 0.236 & $9.21^{+0.05}_{-0.03} \times 10^{-10}$ & $7.90^{+0.05}_{-0.03} \times 10^{-10}$ & $7.64^{+0.04}_{-0.02}$ & $(1.77 \sim 1.81) \times 10^{12}$
       \\
       $\psi(1D) D^{*}$ & 0.00864 & 0.0159 & $9.47^{+0.24}_{-0.03} \times 10^{-6}$ & $9.32^{+0.24}_{-0.03} \times 10^{-6}$ & $0.801^{+0.020}_{-0.030}$ & $(1.38 \sim 1.61) \times 10^{10}$
       \\
       $\psi(1D) D_{s}^{}$ & -0.000229 & -0.0000605 & $2.96^{+0.03}_{-0.04} \times 10^{-4}$ & $2.96^{+0.03}_{-0.04} \times 10^{-4}$ & $-0.0213^{+0.0002}_{-0.0002}$ & $(6.51 \sim 6.98) \times 10^{12}$
       \\
       $\psi(1D) D_{s 0}^{*}$ & -0.00146 & -0.00418 & $2.54^{+0.05}_{-0.06} \times 10^{-7}$ & $2.55^{+0.05}_{-0.06} \times 10^{-7}$ & $-0.136^{+0.005}_{-0.004}$ & $(1.74 \sim 2.07) \times 10^{12}$
       \\
       $\psi(1D) D_{s}^{*}$ & -0.000291 & -0.000529 & $1.70^{+0.03}_{-0.02} \times 10^{-4} $ & $1.71^{+0.03}_{-0.02} \times 10^{-4}$ & $-0.0271^{+0.0005}_{-0.0003}$ & $(6.81 \sim  7.44) \times 10^{11}$
       \\
       $\psi(1D) D_{s 1}^{}$ & 0.000141 & 0.000290 & $2.48^{+0.17}_{-0.19}\times 10^{-4}$ & $2.48^{+0.17}_{-0.19} \times 10^{-4}$ & $-0.0131^{+0.0014}_{-0.0016}$ & $(1.61 \sim 2.98) \times 10^{13}$
       \\
       \colrule
       $\psi(3S) D^{}$ & -0.00783 & -0.00114 & $3.62^{+0.04}_{-0.02} \times 10^{-8}$ & $3.67^{+0.04}_{-0.02} \times 10^{-8}$ & $-0.731^{+0.003}_{-0.008}$ & $(4.53 \sim 4.75) \times 10^{12}$
       \\
       $\psi(3S) D^{*}$ & 0.00559 & 0.0109 & $3.38^{+0.12}_{-0.09} \times 10^{-5}$ & $3.34^{+0.12}_{-0.09} \times 10^{-5}$ & $0.519^{+0.021}_{-0.018}$ & $(8.86 \sim 10.97) \times 10^{9}$
       \\
       $\psi(3S) D_{s}^{}$ & -0.0000571 & -0.0000153 & $3.76^{+0.49}_{-0.66} \times 10^{-5}$ & $3.76^{+0.49}_{-0.66} \times 10^{-5}$ & $-0.00533^{+0.00020}_{-0.00020}$ & $(7.43 \sim 10.116) \times 10^{13}$
       \\
       $\psi(3S) D_{s}^{*}$ & -0.000277 & -0.000545 & {$5.89^{+0.21}_{-0.87} \times 10^{-4} $} & {$5.90^{+0.20}_{-0.38} \times 10^{-4}$} & $-0.0258^{+0.0012}_{-0.0023}$ & $(1.51 \sim 5.63) \times 10^{11}$
       \\
    \end{tabular}
  \end{ruledtabular}
\end{table}

\begin{figure}[htb]
  \centering
  \subfigure[]{\includegraphics[scale=0.52]{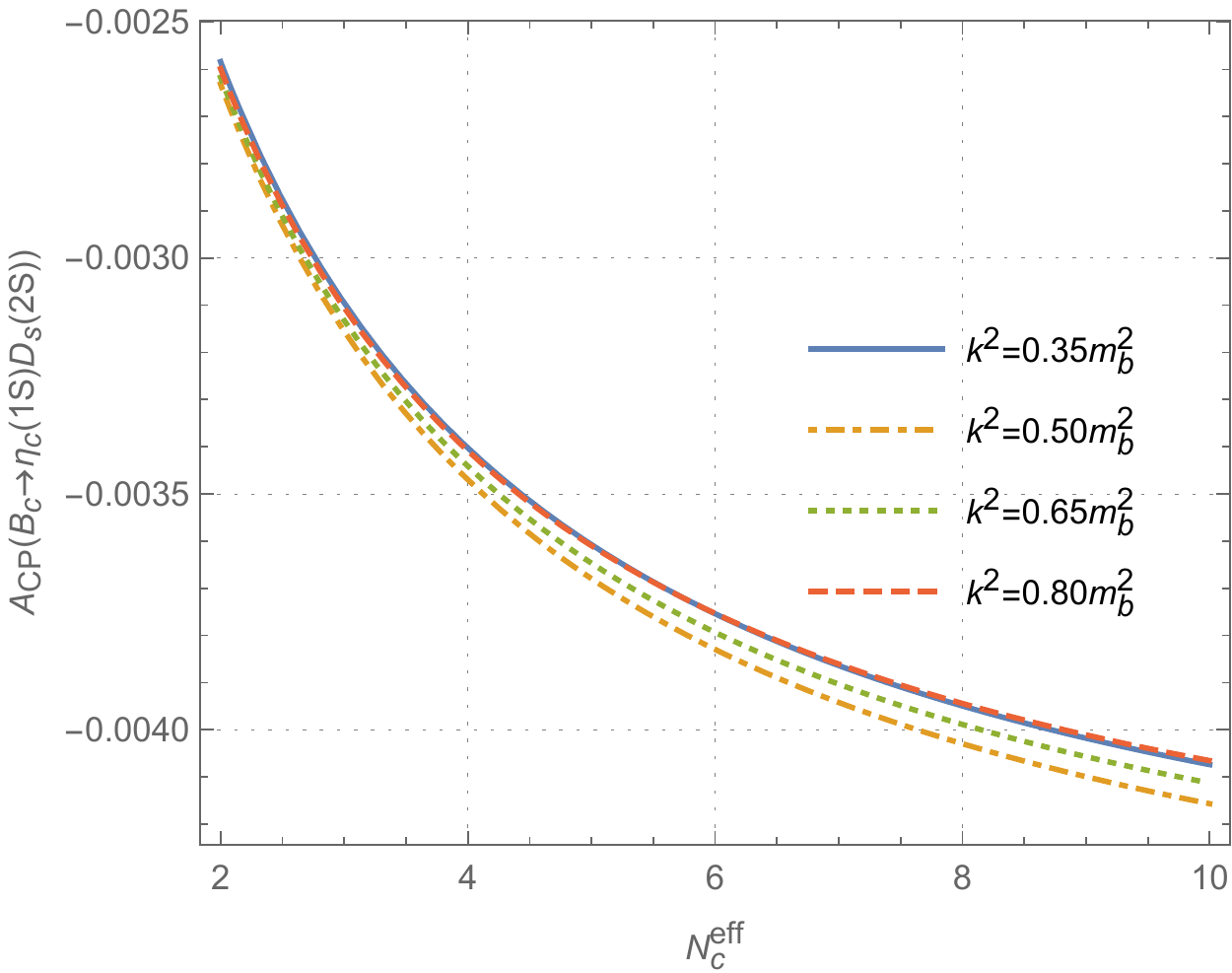}}
  \hspace{0.8cm}
  \subfigure[]{\includegraphics[scale=0.51]{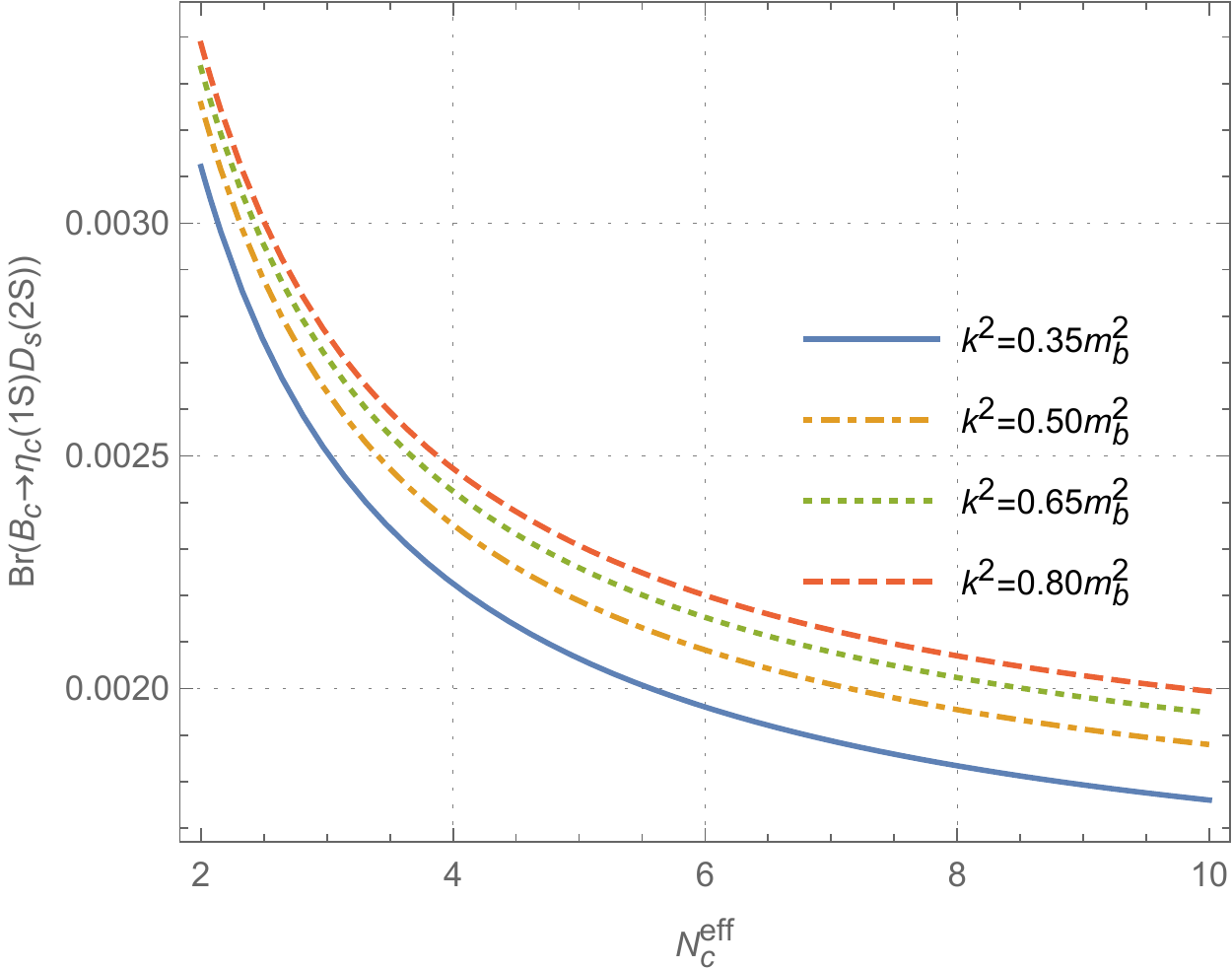}}\\
  \caption[]{\label{fig: bcu2sD1s} $\mathcal A_{cp}$ and averaged branching ratio of $B_c \rightarrow \psi(1S)D(2S)$ change wiht $k^2$ and $N_c^{eff}$. }
\end{figure}

\begin{figure}[htb]
  \centering
  \subfigure[]{\includegraphics[scale=0.48]{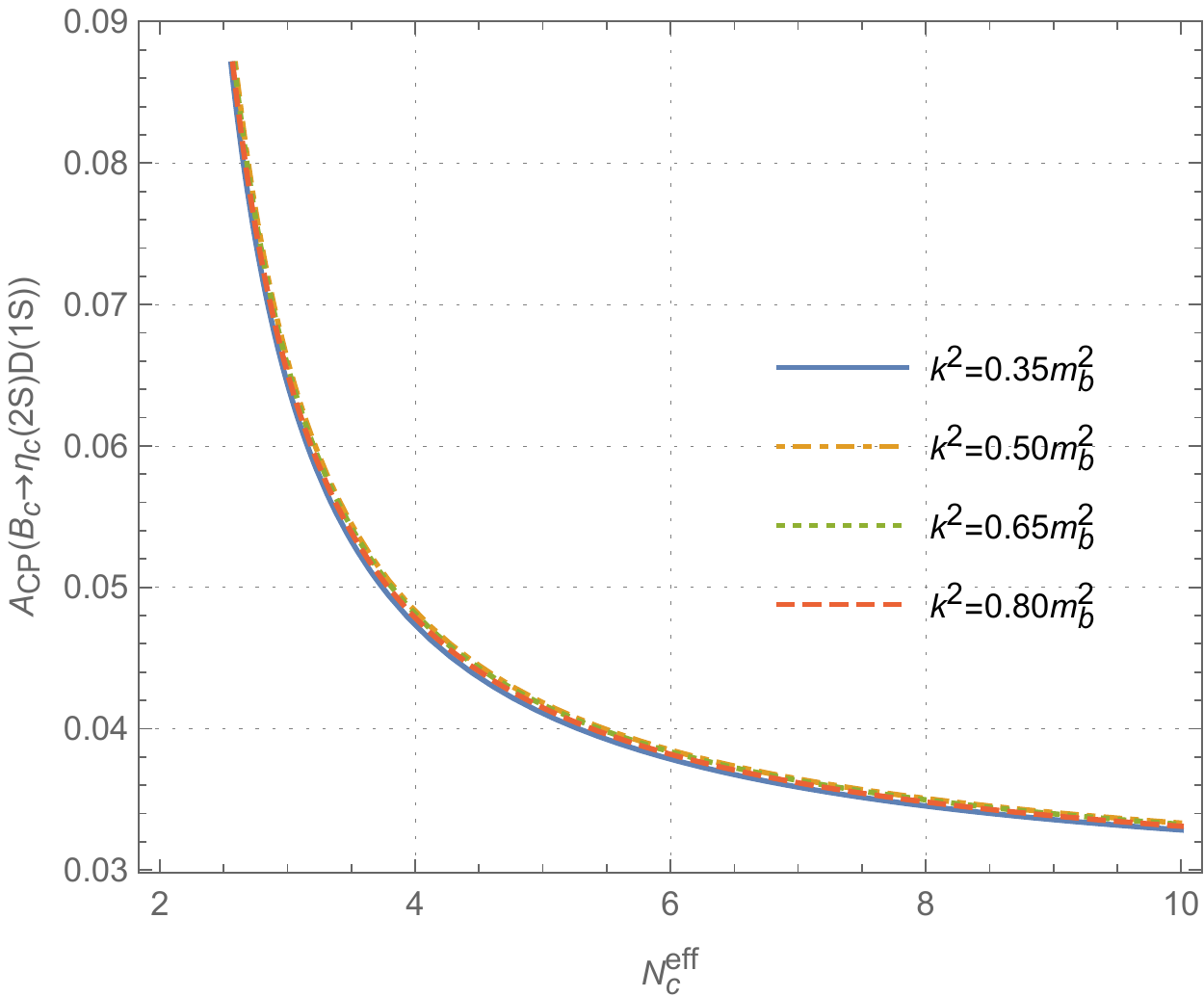}}
  \hspace{0.7cm}
  \subfigure[]{\includegraphics[scale=0.52]{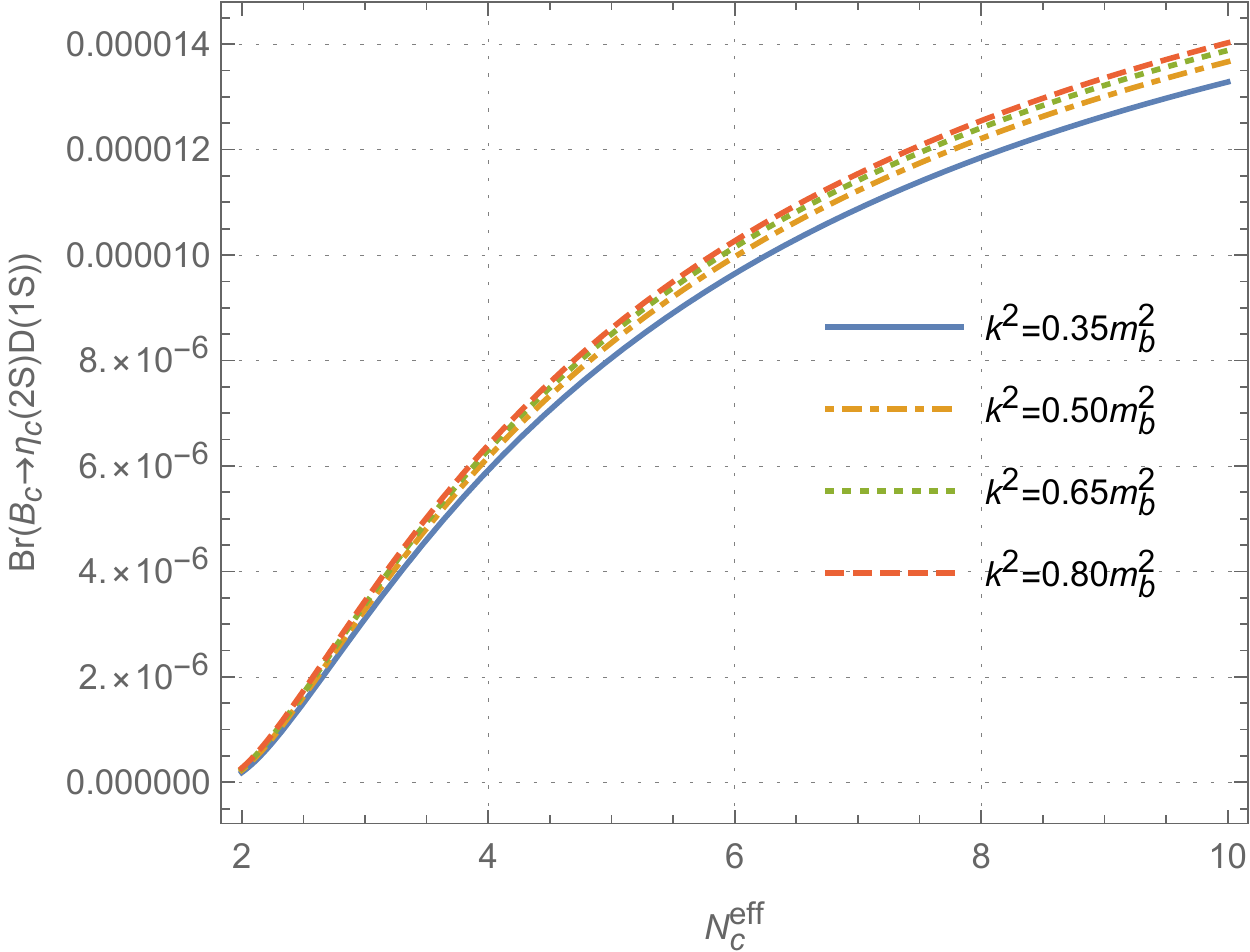}}\\
  \caption[]{\label{fig: bcnc1sD2s} $\mathcal A_{cp}$ and averaged branching ratio of $B_c \rightarrow \eta_c(2S)D(1S)$ change with $k^2$ and $N_c^{eff}$. }
\end{figure}

\begin{figure}[htb]
  \centering
  \subfigure[]{\includegraphics[scale=0.5]{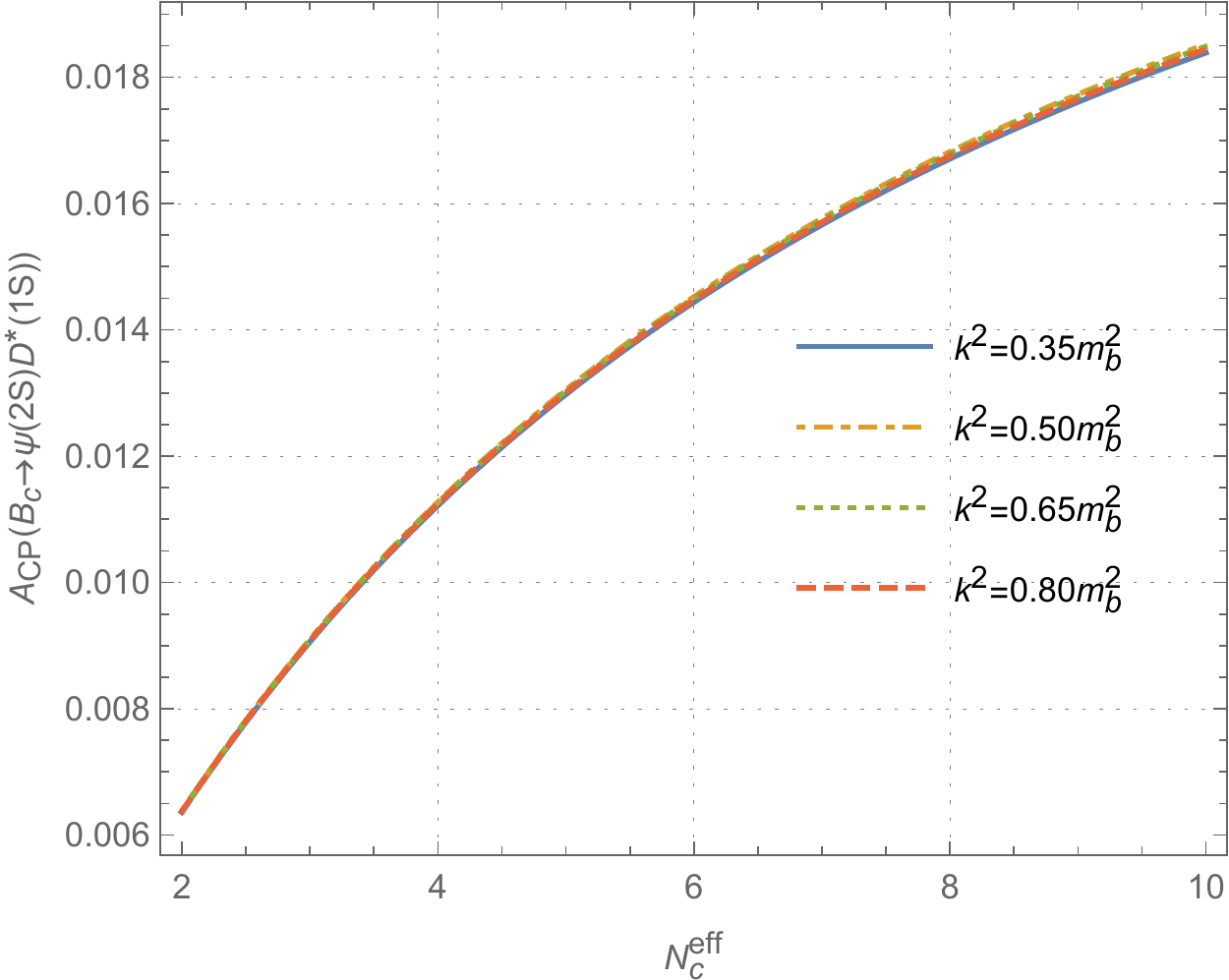}}
  \hspace{0.8cm}
  \subfigure[]{\includegraphics[scale=0.52]{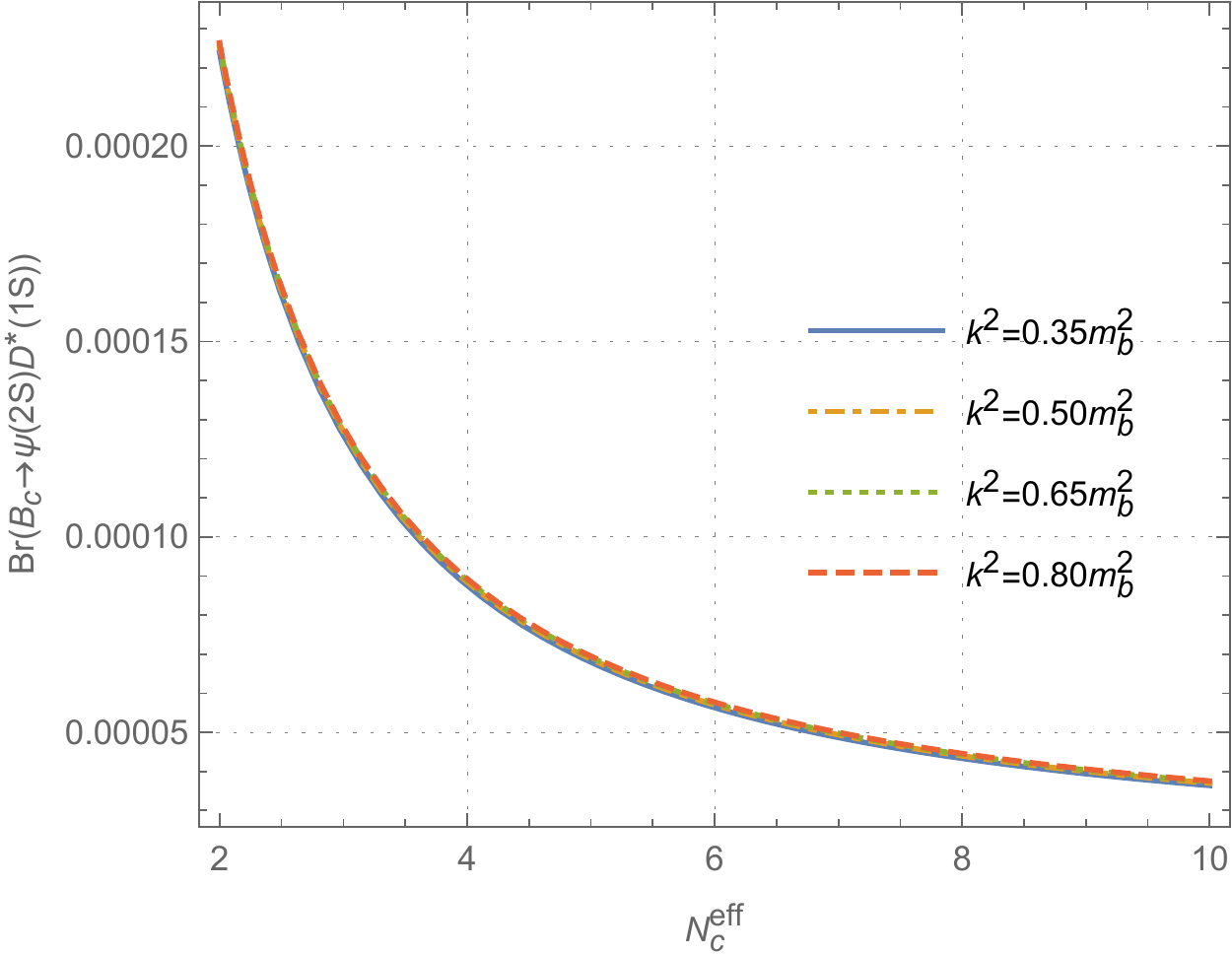}}\\
  \caption[]{\label{fig: bcnc2sD1P0+} $\mathcal A_{cp}$ and averaged branching ratio of $B_c \rightarrow \psi(2S)D^*(1S)$ change with $k^2$ and $N_c^{eff}$. }
\end{figure}

\begin{figure}[htb]
  \centering
  \subfigure[]{\includegraphics[scale=0.49]{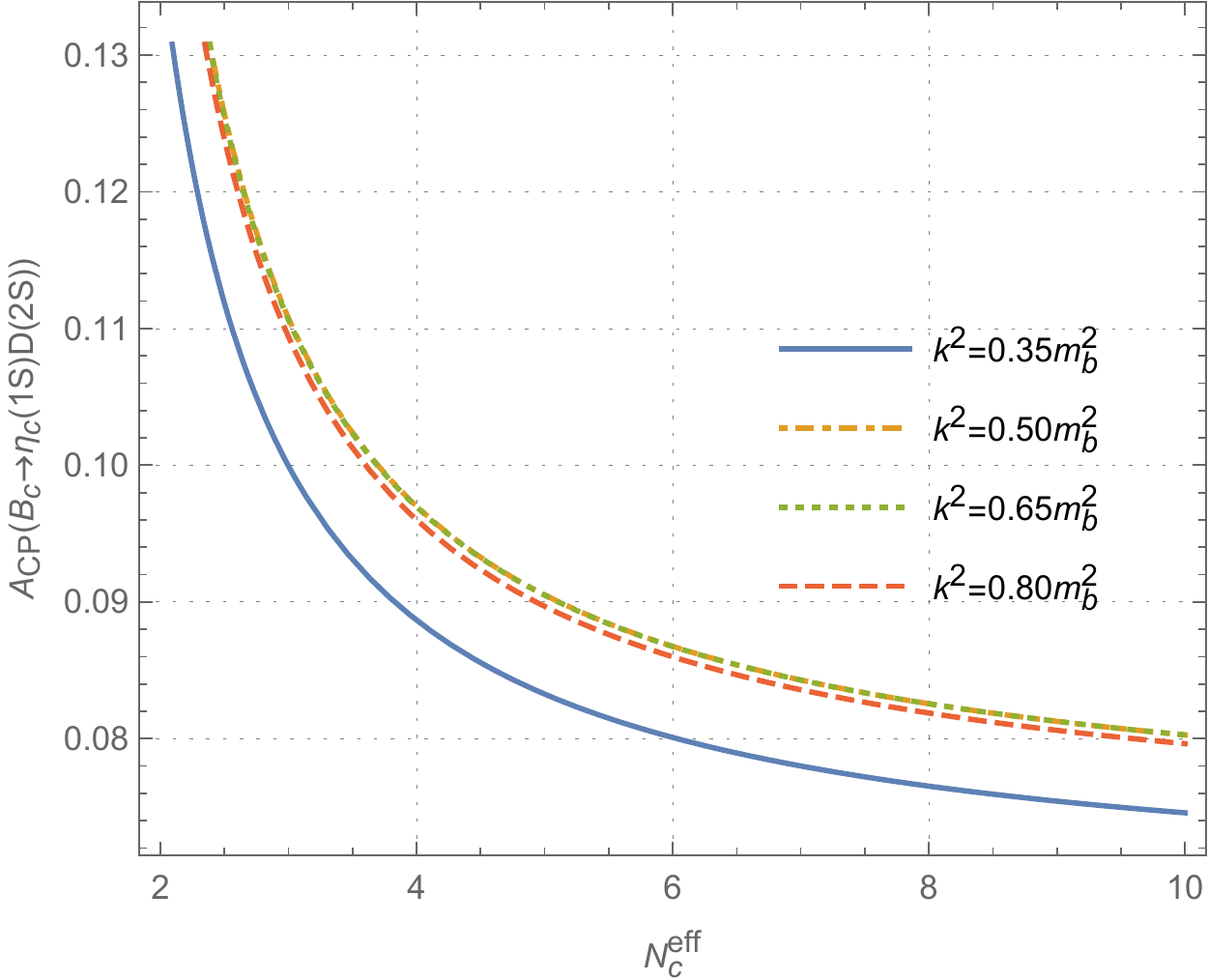}}
  \hspace{0.8cm}
  \subfigure[]{\includegraphics[scale=0.52]{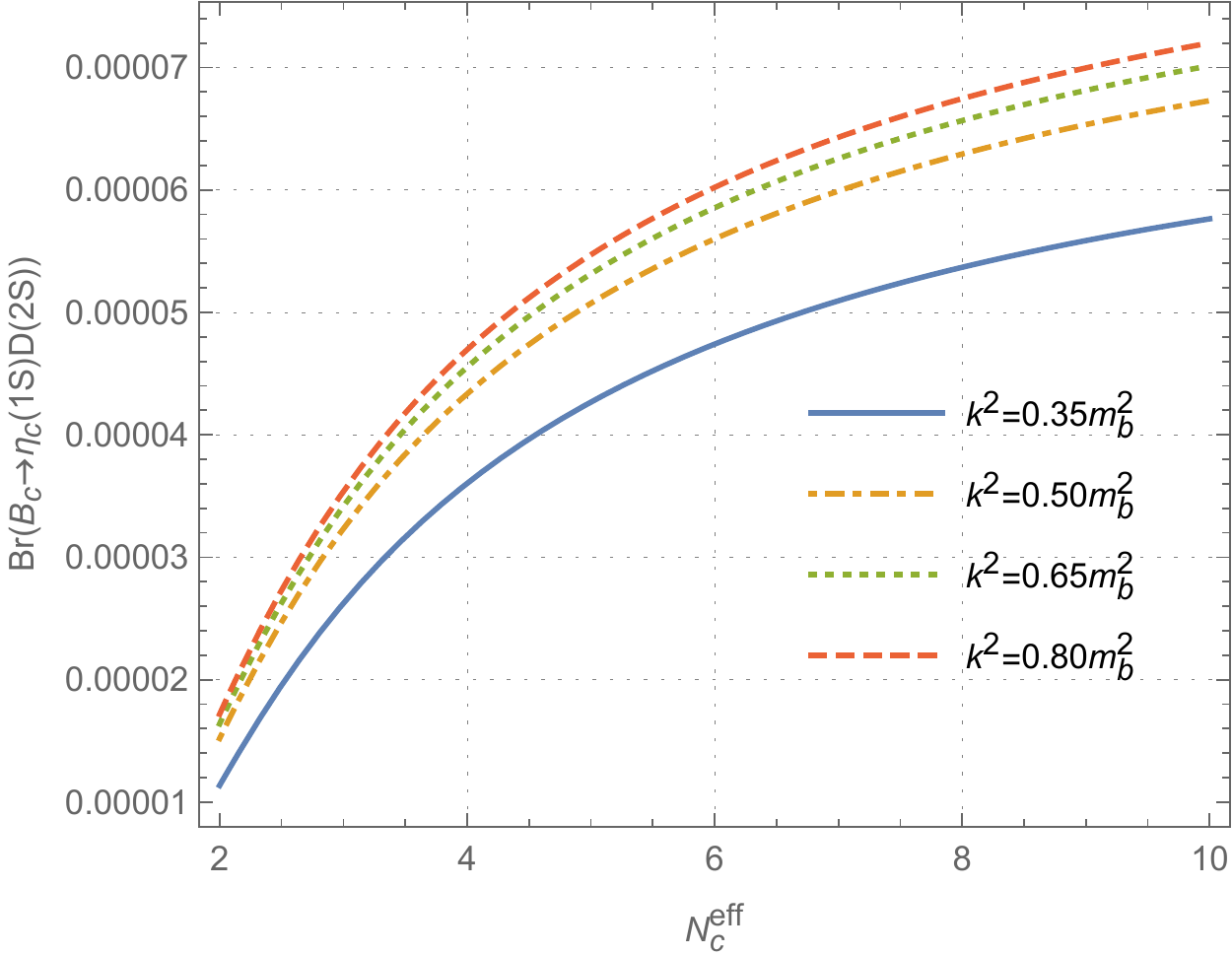}}\\
  \caption[]{\label{fig: bcnc2sD1s} $\mathcal A_{cp}$ and averaged branching ratio of $B_c \rightarrow \eta_c(1S)D(2S)$ change with $k^2$ and $N_c^{eff}$. }
\end{figure}

\begin{figure}[htb]
  \centering
  \subfigure[]{\includegraphics[scale=0.5]{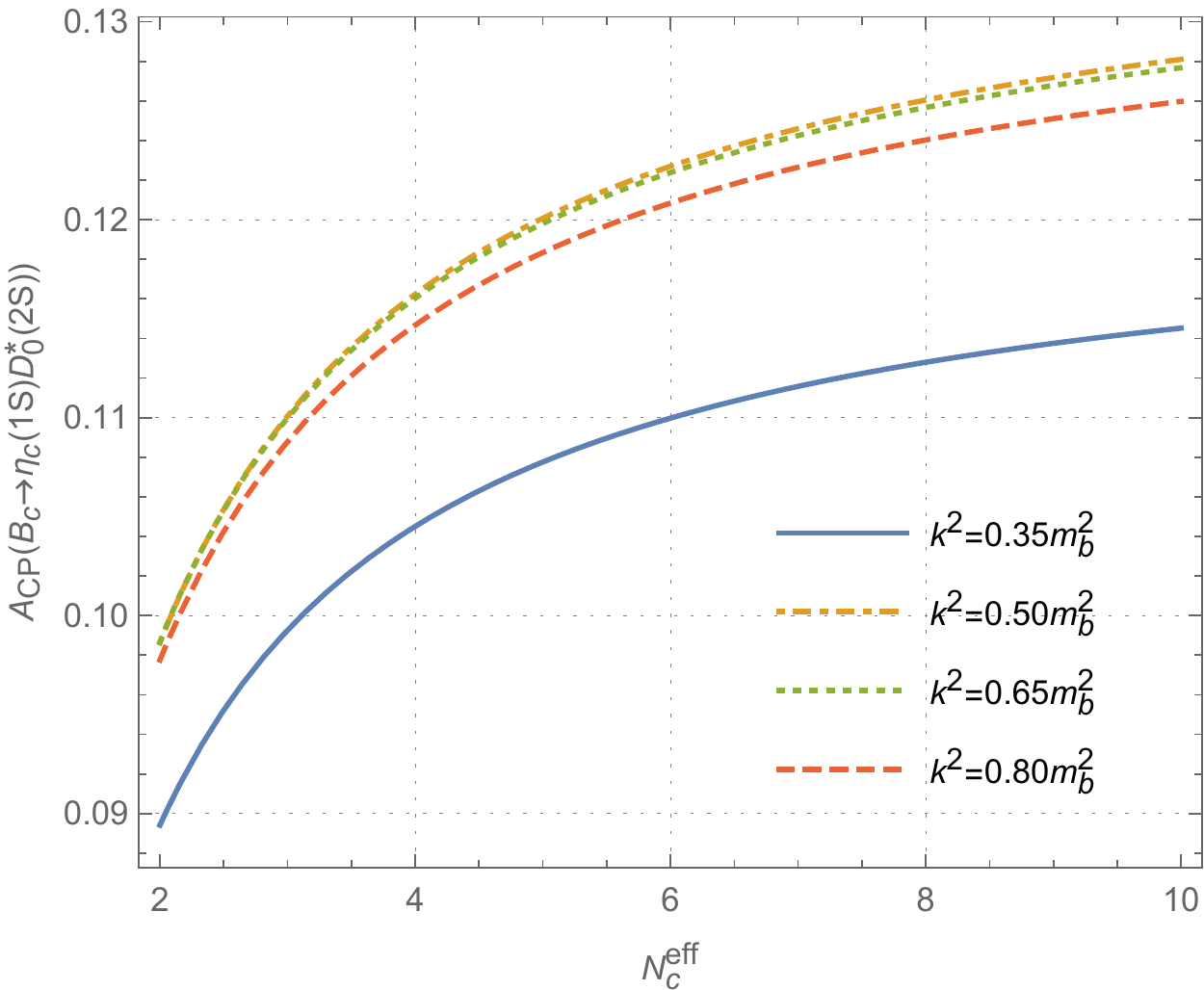}}
  \hspace{0.8cm}
  \subfigure[]{\includegraphics[scale=0.53]{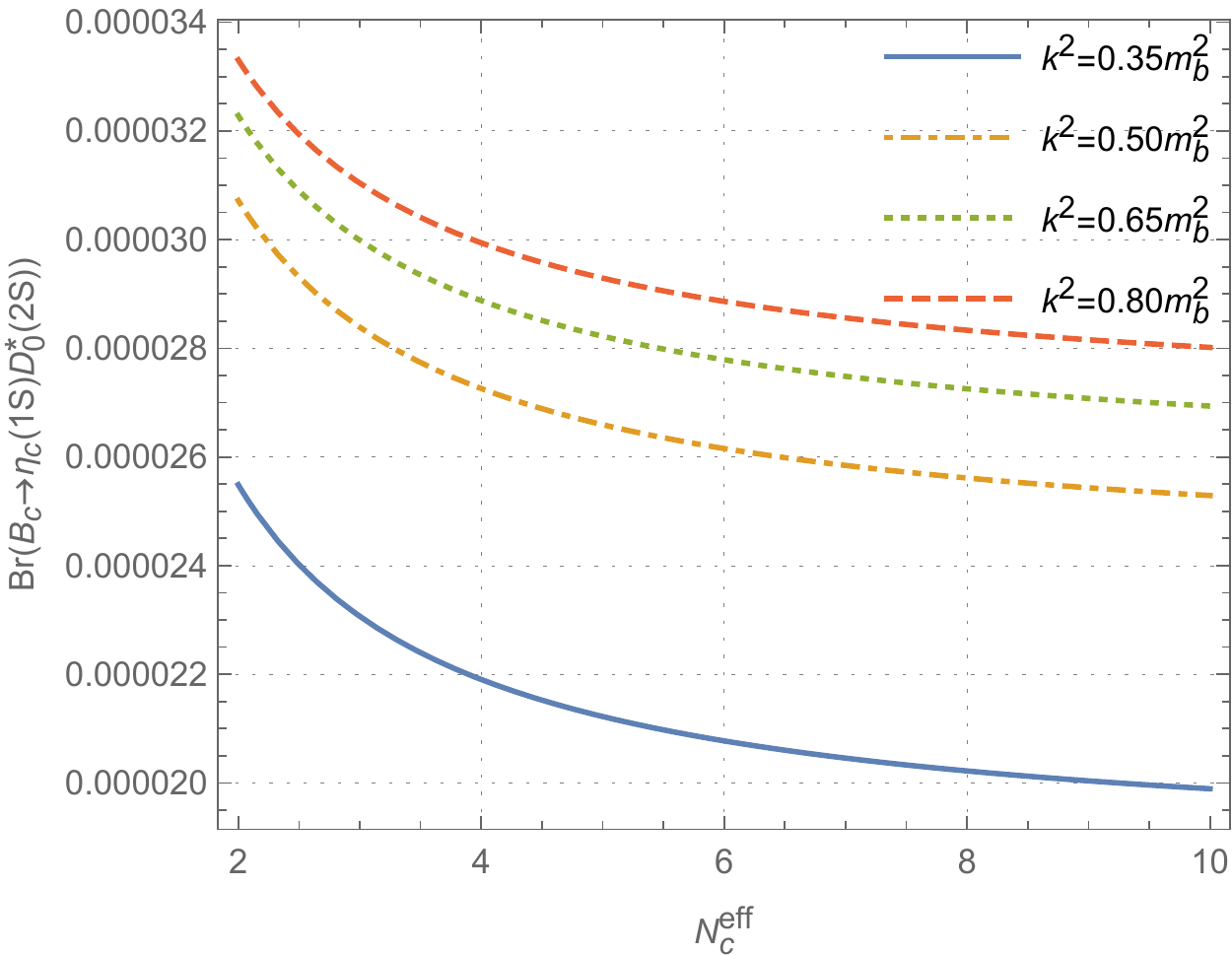}}\\
  \caption[]{\label{fig: bcnc2sD1sstar} $\mathcal A_{cp}$ and averaged branching ratio of $B_c \rightarrow \eta_c(2S)D_0^*(1S)$ change with $k^2$ and $N_c^{eff}$. }
\end{figure}

\begin{figure}[htb]
  \centering
  \subfigure[]{\includegraphics[scale=0.51]{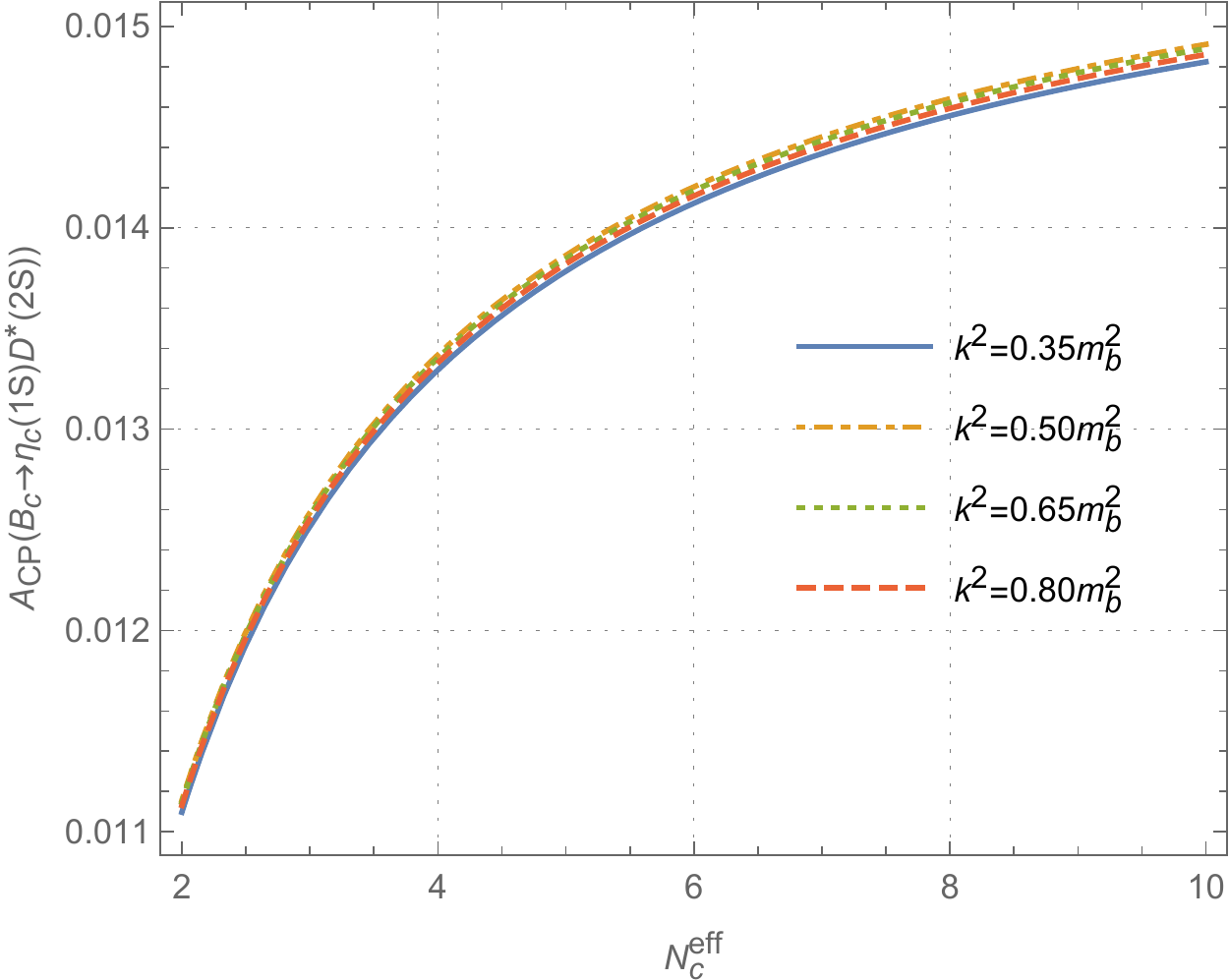}}
  \hspace{0.8cm}
  \subfigure[]{\includegraphics[scale=0.53]{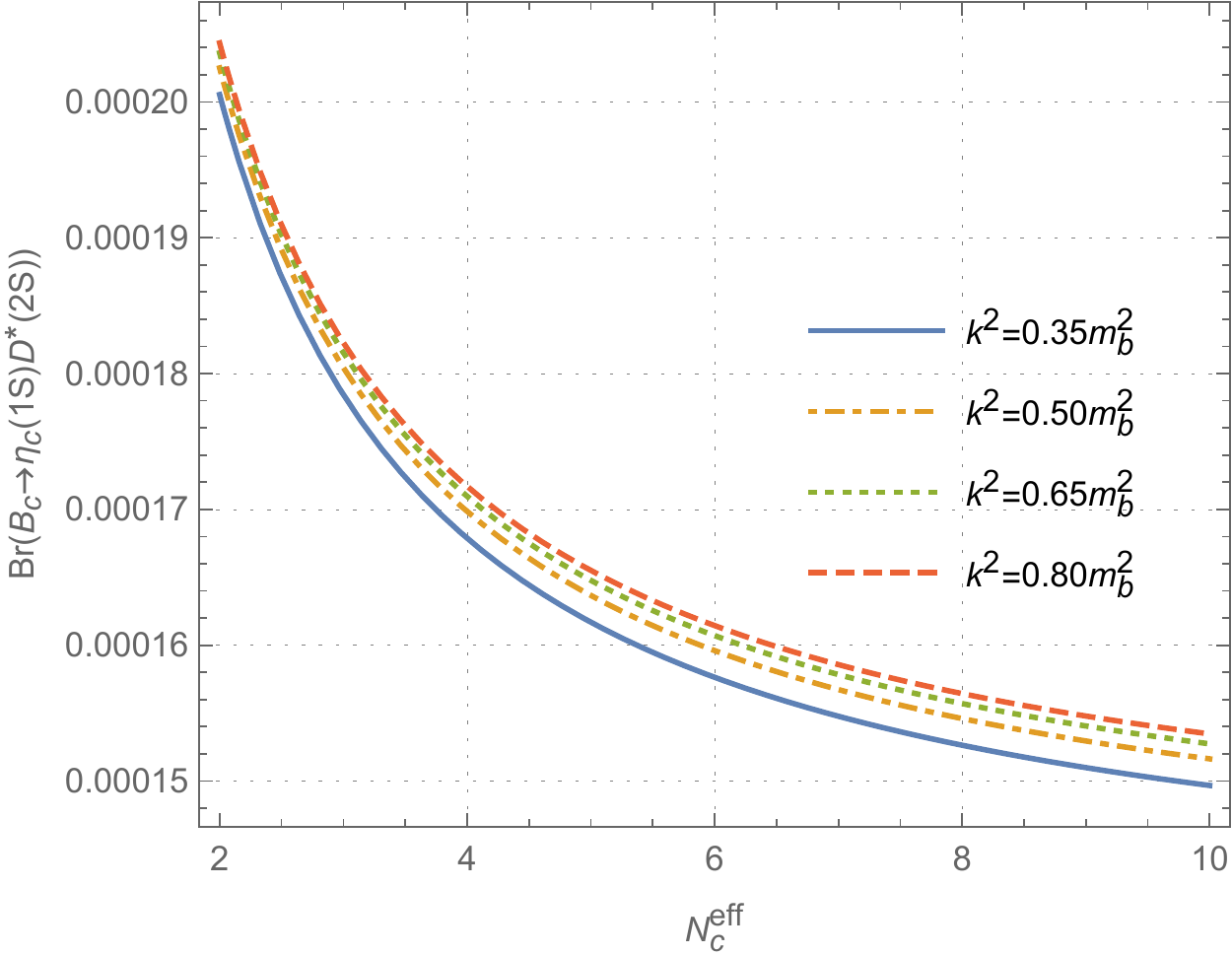}}\\
  \caption[]{\label{fig: bcnc2sDs1s} $\mathcal A_{cp}$ and averaged branching ratio of $B_c \rightarrow \eta_c(1S)D^*(2S)$ change with $k^2$ and $N_c^{eff}$. }
\end{figure}

\begin{figure}[htb]
  \centering
  \subfigure[]{\includegraphics[scale=0.5]{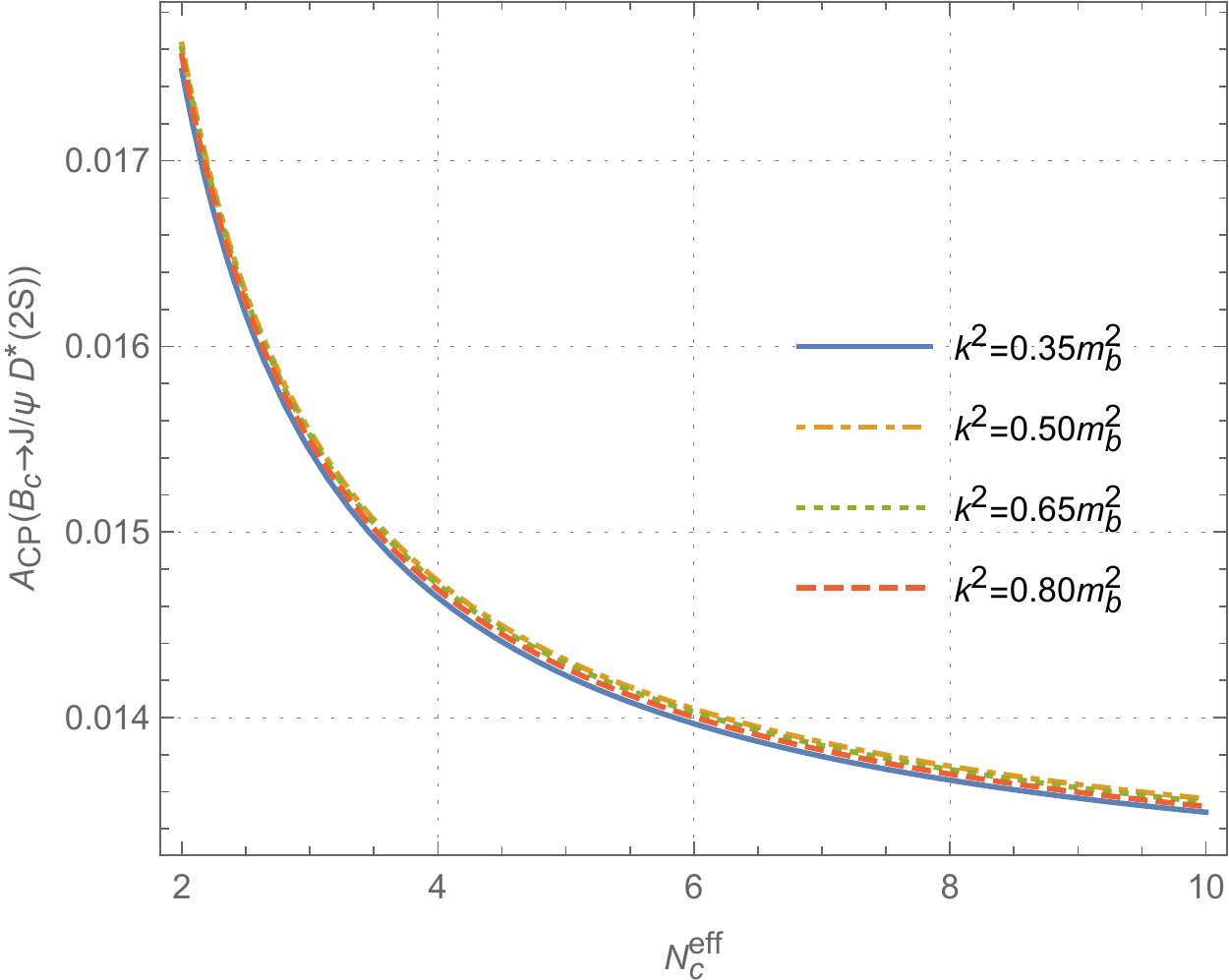}}
  \hspace{0.8cm}
  \subfigure[]{\includegraphics[scale=0.51]{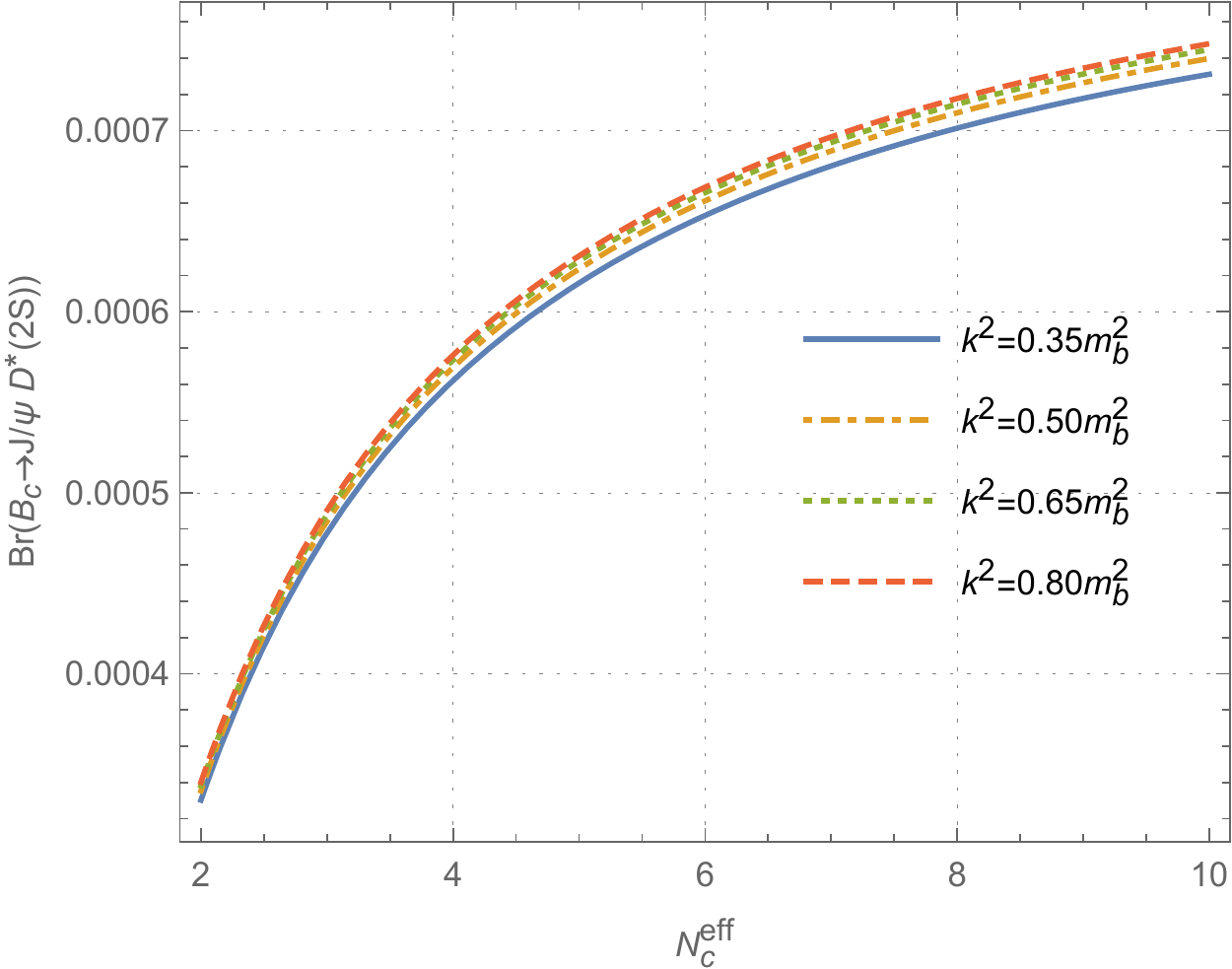}}\\
  \caption[]{\label{fig: bcu2ssD1sstar} $\mathcal A_{cp}$ and averaged branching ratio of $B_c \rightarrow J/\psi D^*(2S)$ change with $k^2$ and $N_c^{eff}$. }
\end{figure}

\begin{figure}[htb]
 \centering
 \subfigure[]{\includegraphics[scale=0.29]{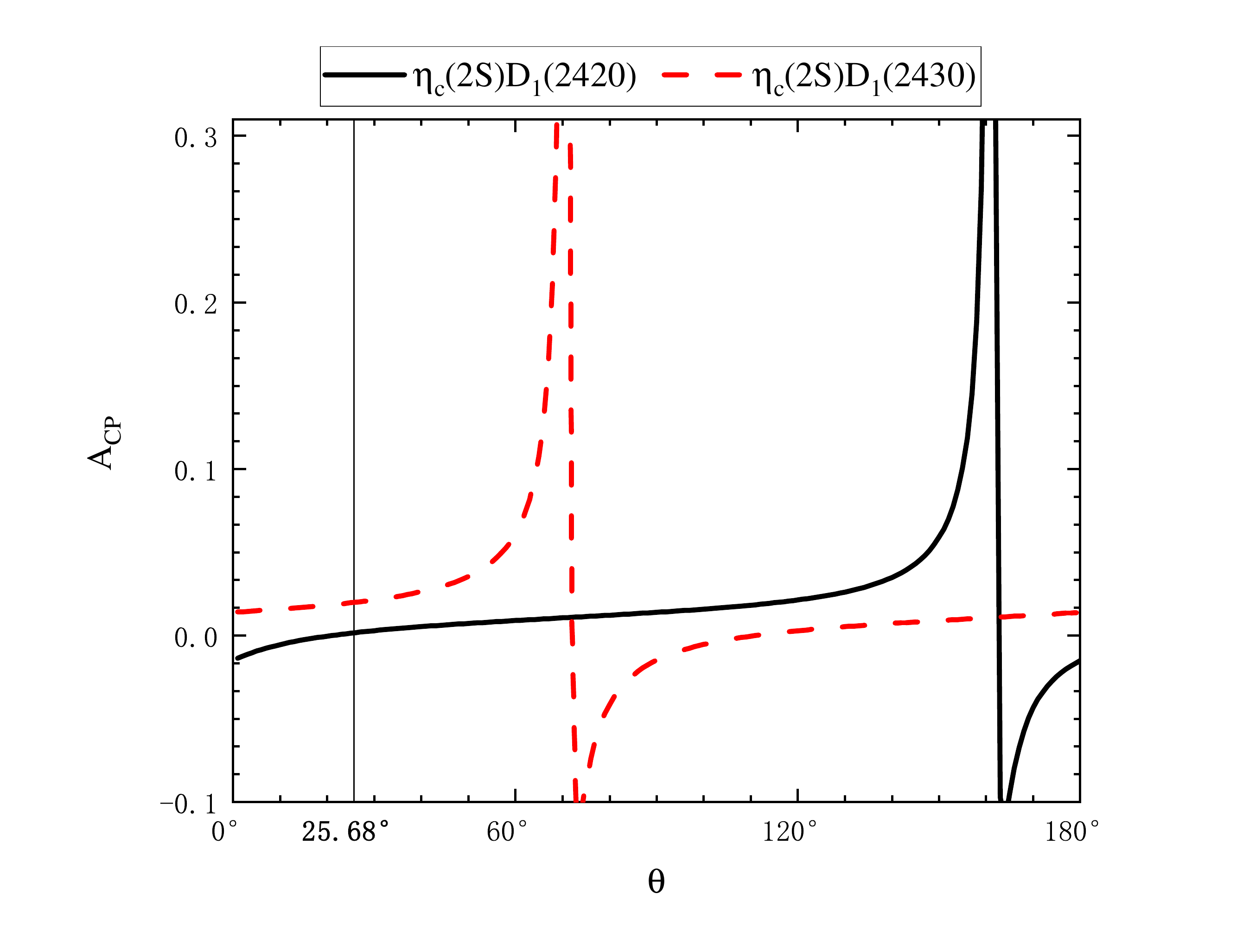}}
 \subfigure[]{\includegraphics[scale=0.29]{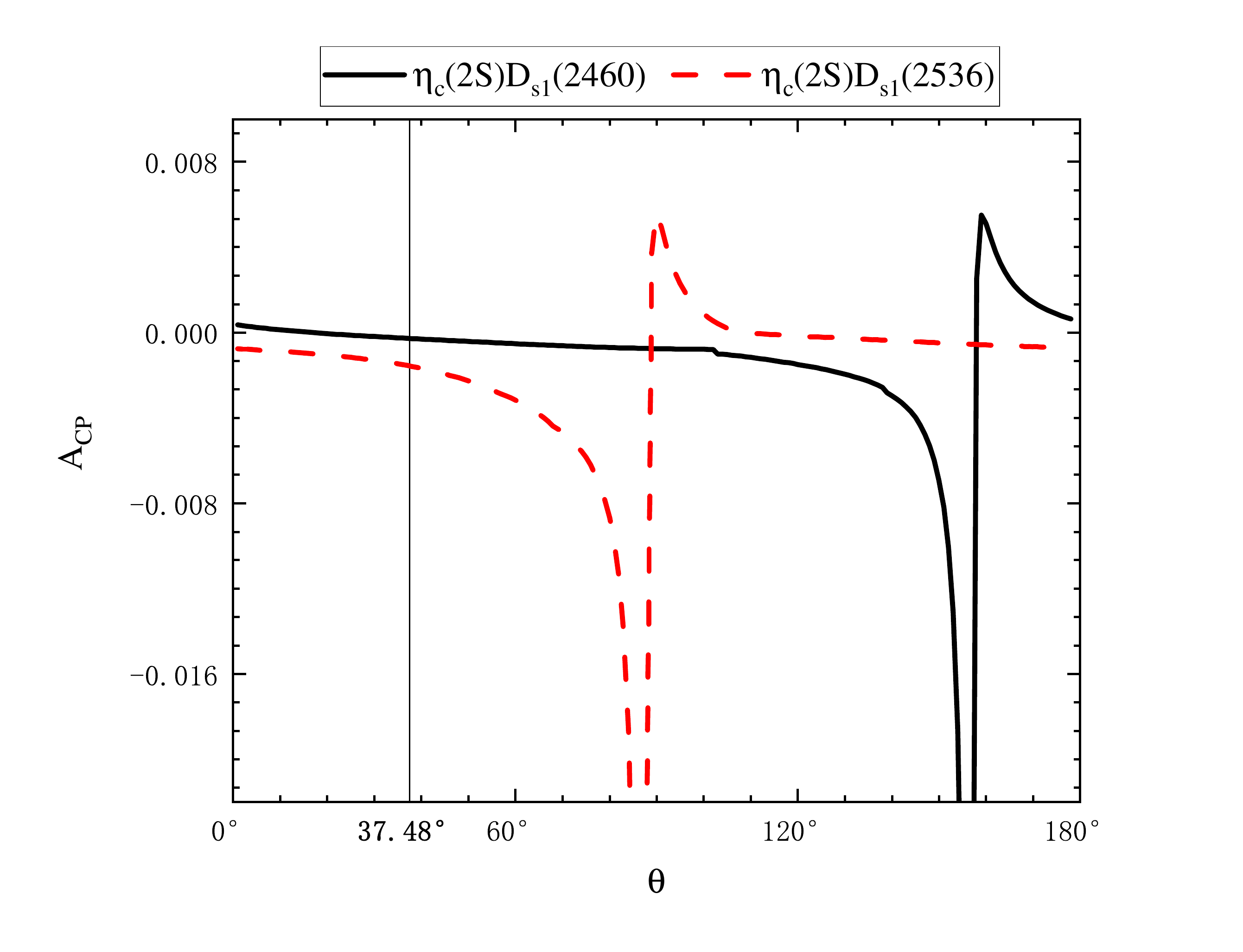}}
 \subfigure[]{\includegraphics[scale=0.29]{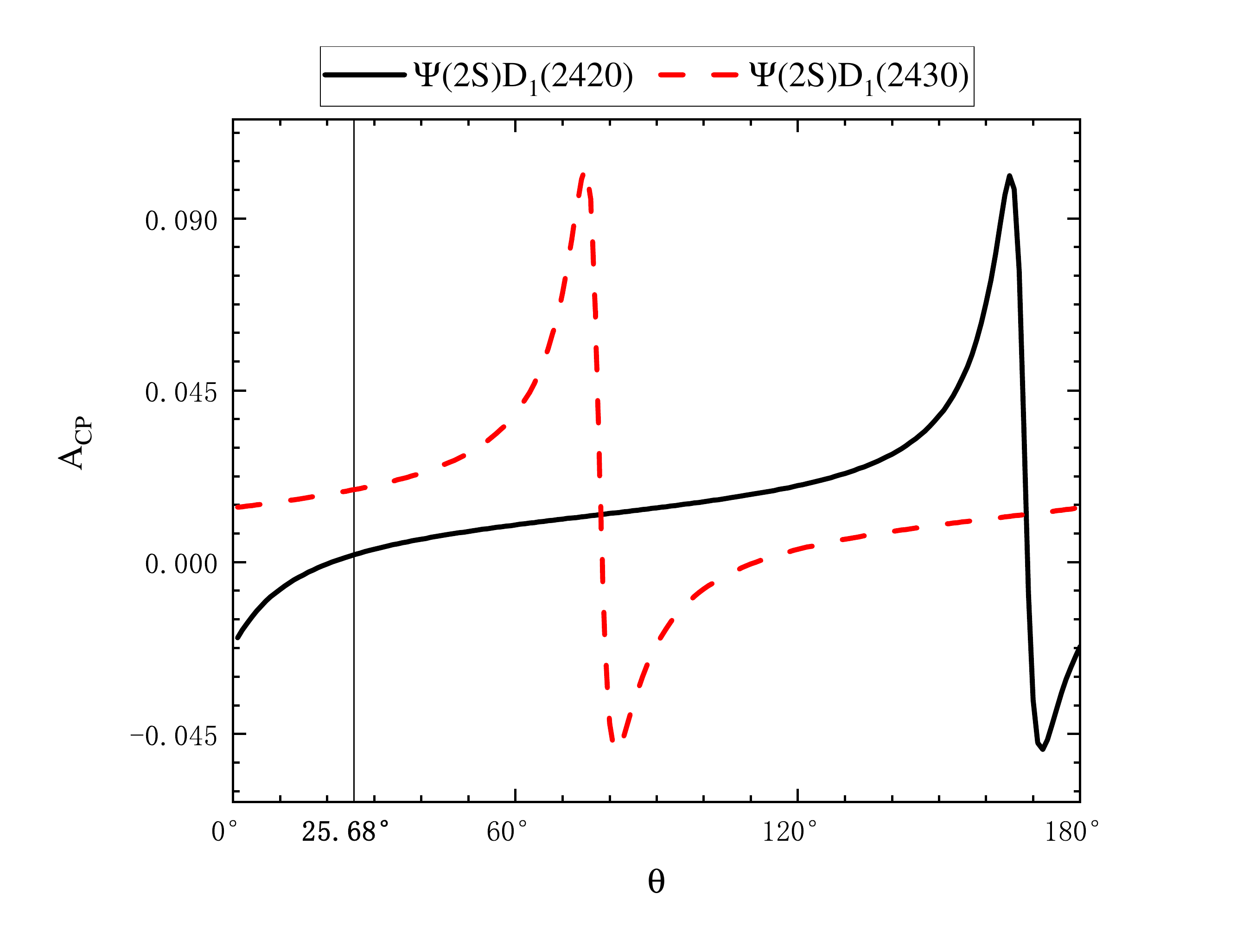}}
 \subfigure[]{\includegraphics[scale=0.29]{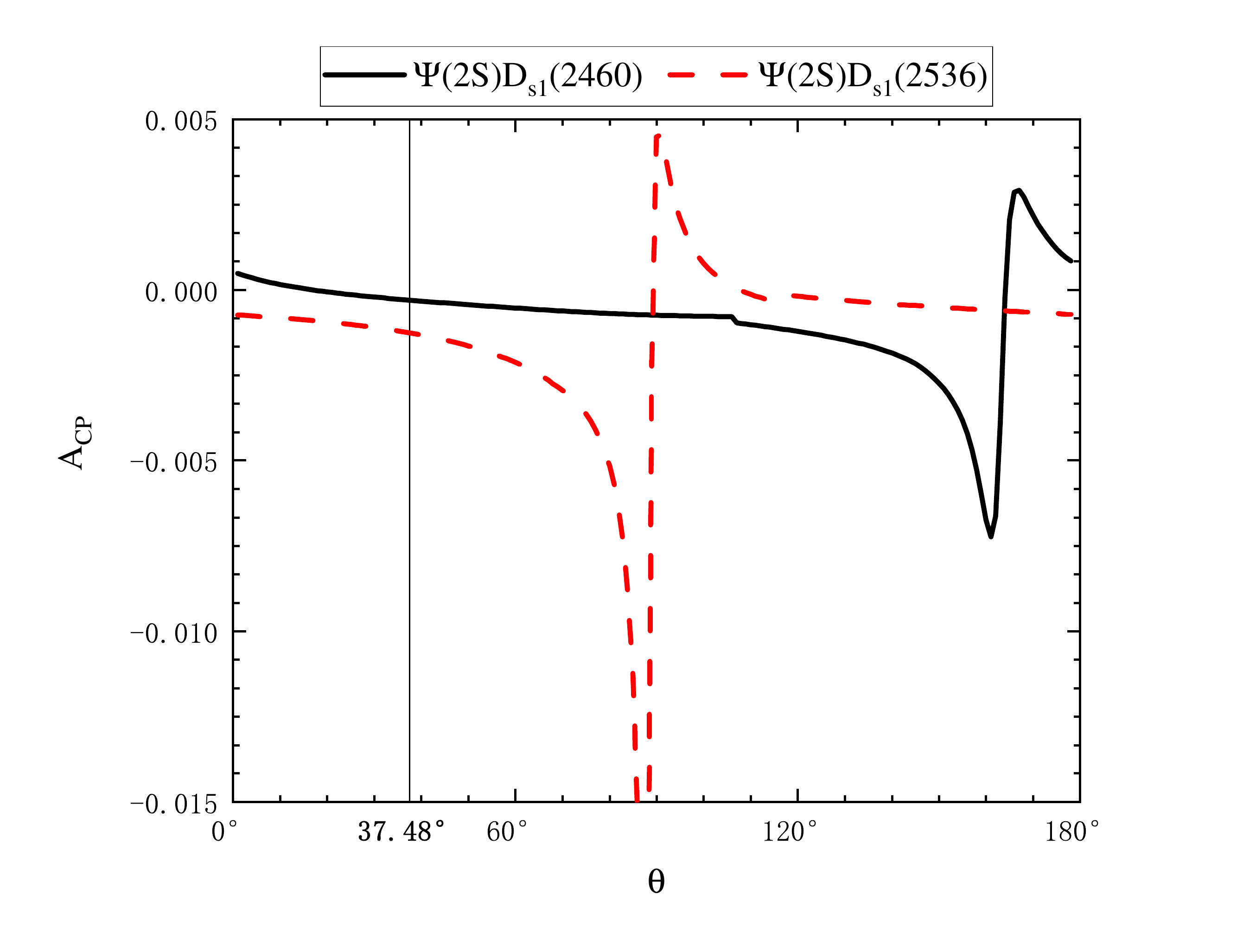}}\\
 \caption[]{$A_{cp}$ varies along with the mixing angle for the $B_c\to c\bar c(2S) + D_{(s)1}(1P)$.}
\end{figure}

\begin{figure}[htb]
 \centering
 \subfigure[]{\includegraphics[scale=0.29]{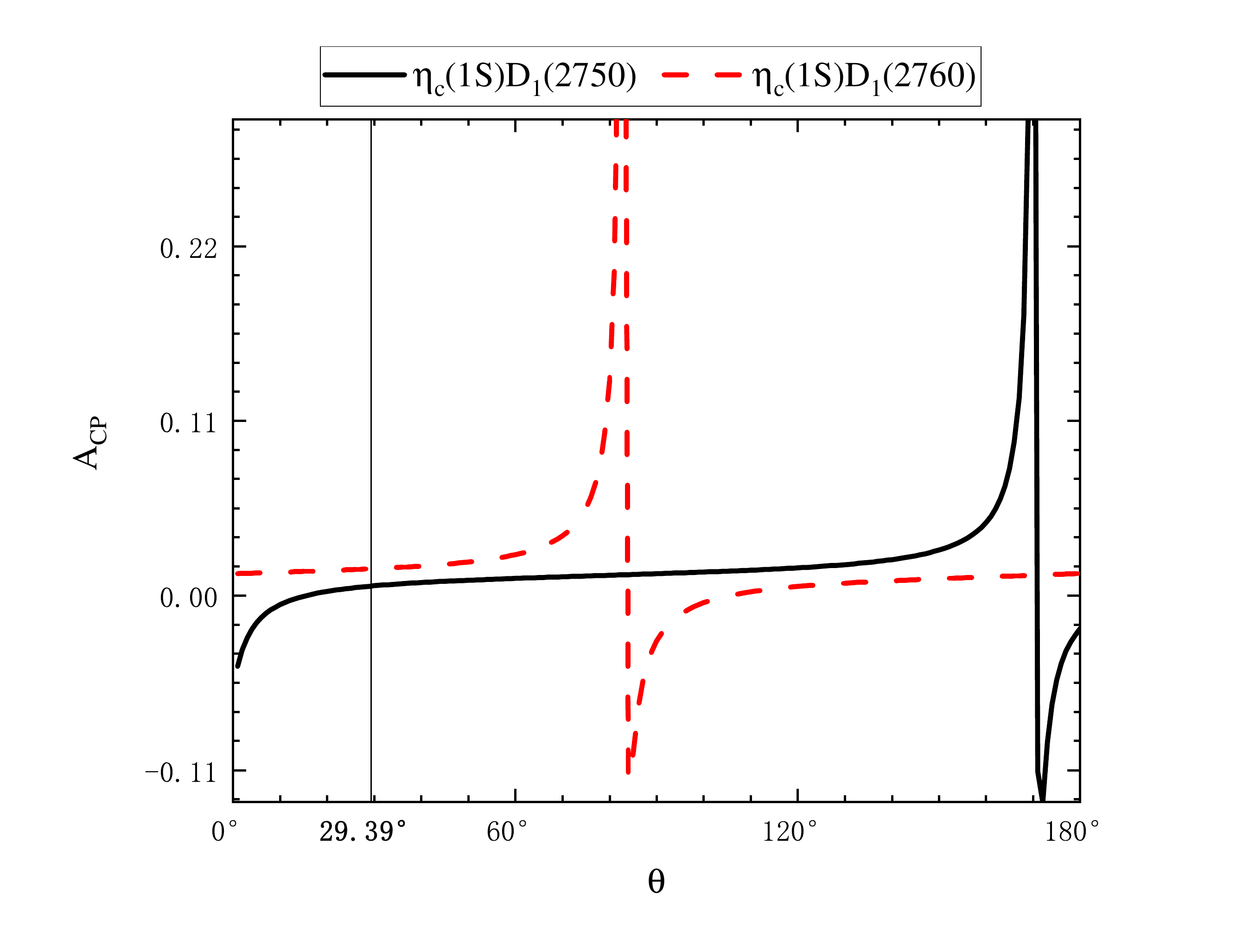}}
 \subfigure[]{\includegraphics[scale=0.29]{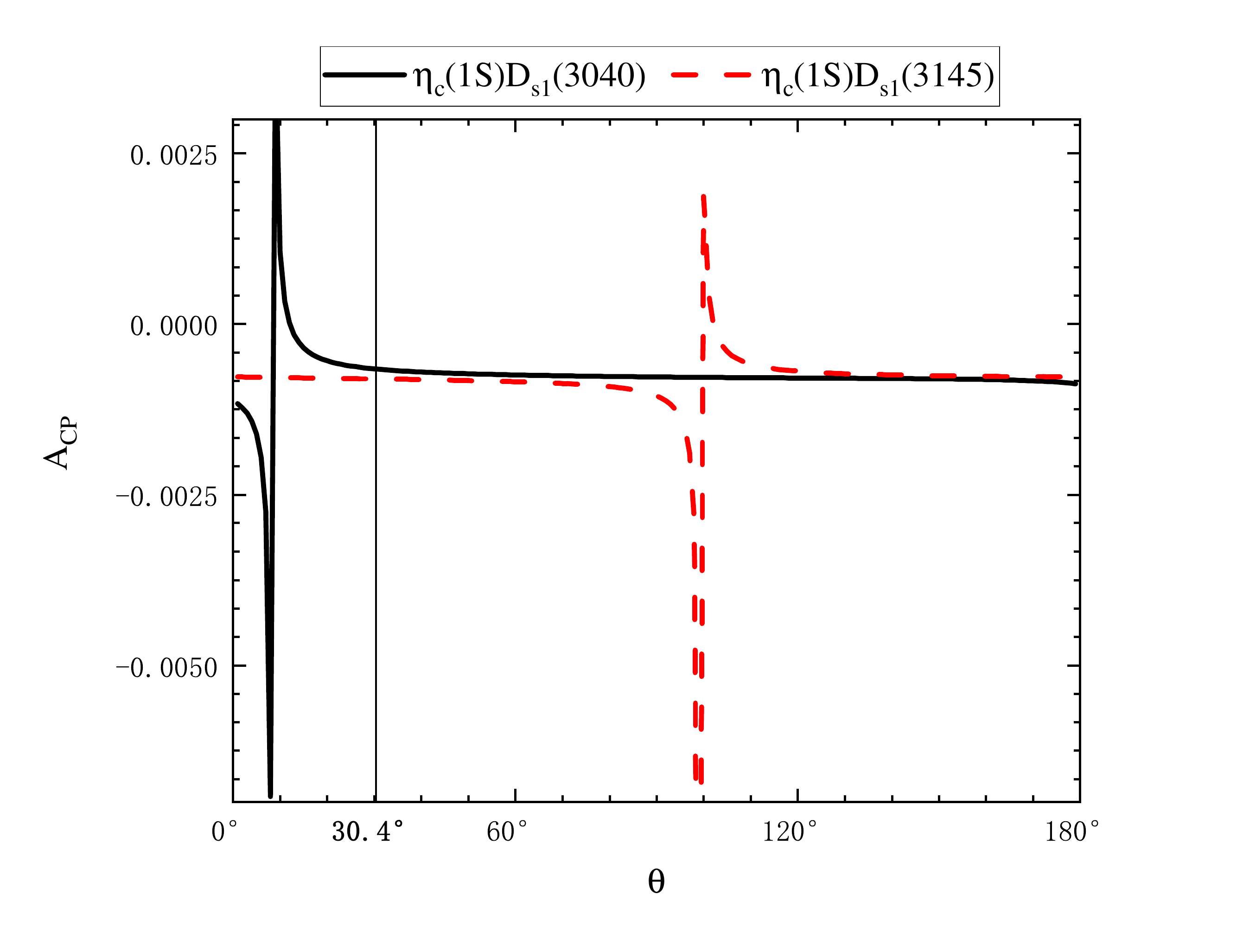}}
 \subfigure[]{\includegraphics[scale=0.29]{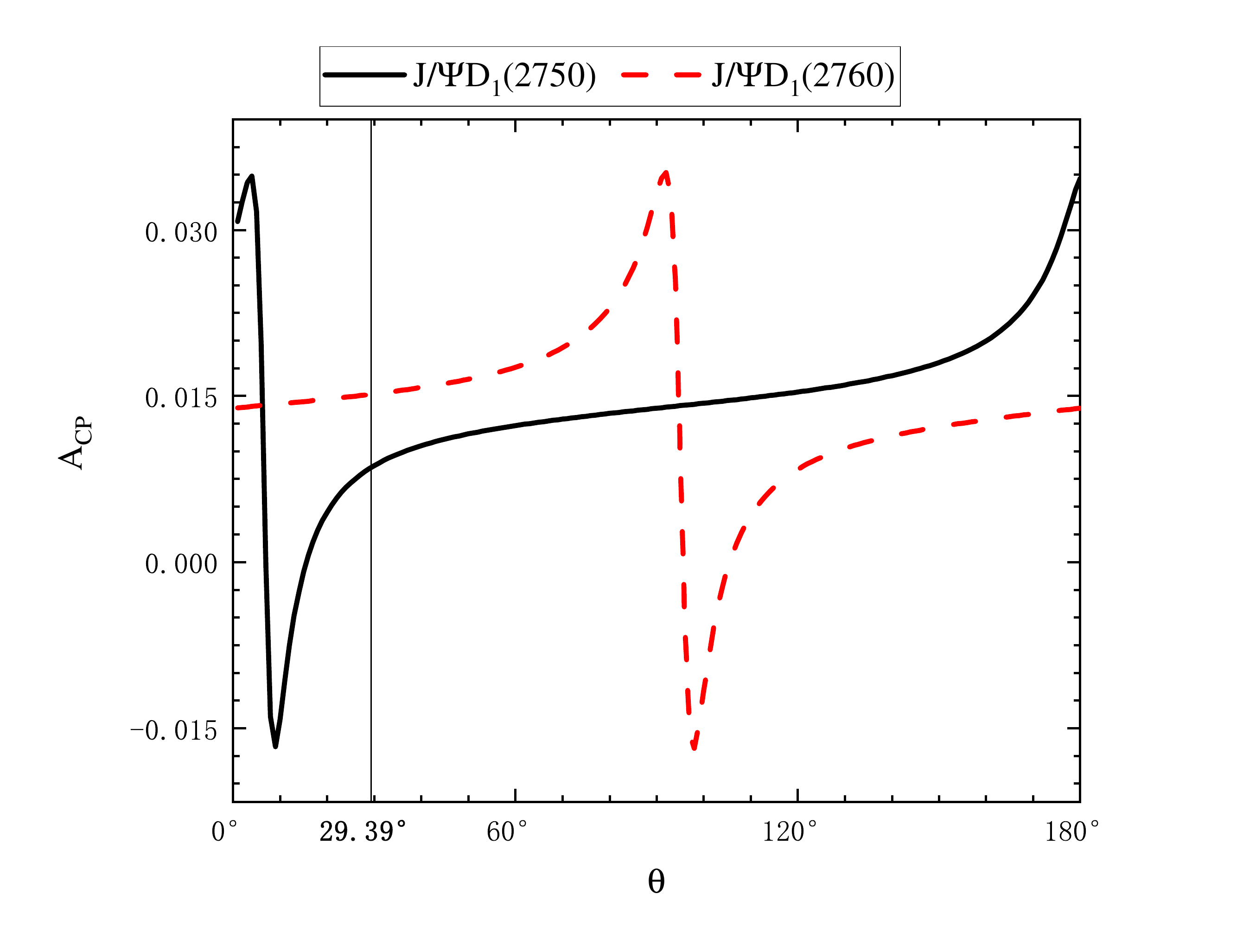}}
 \subfigure[]{\includegraphics[scale=0.29]{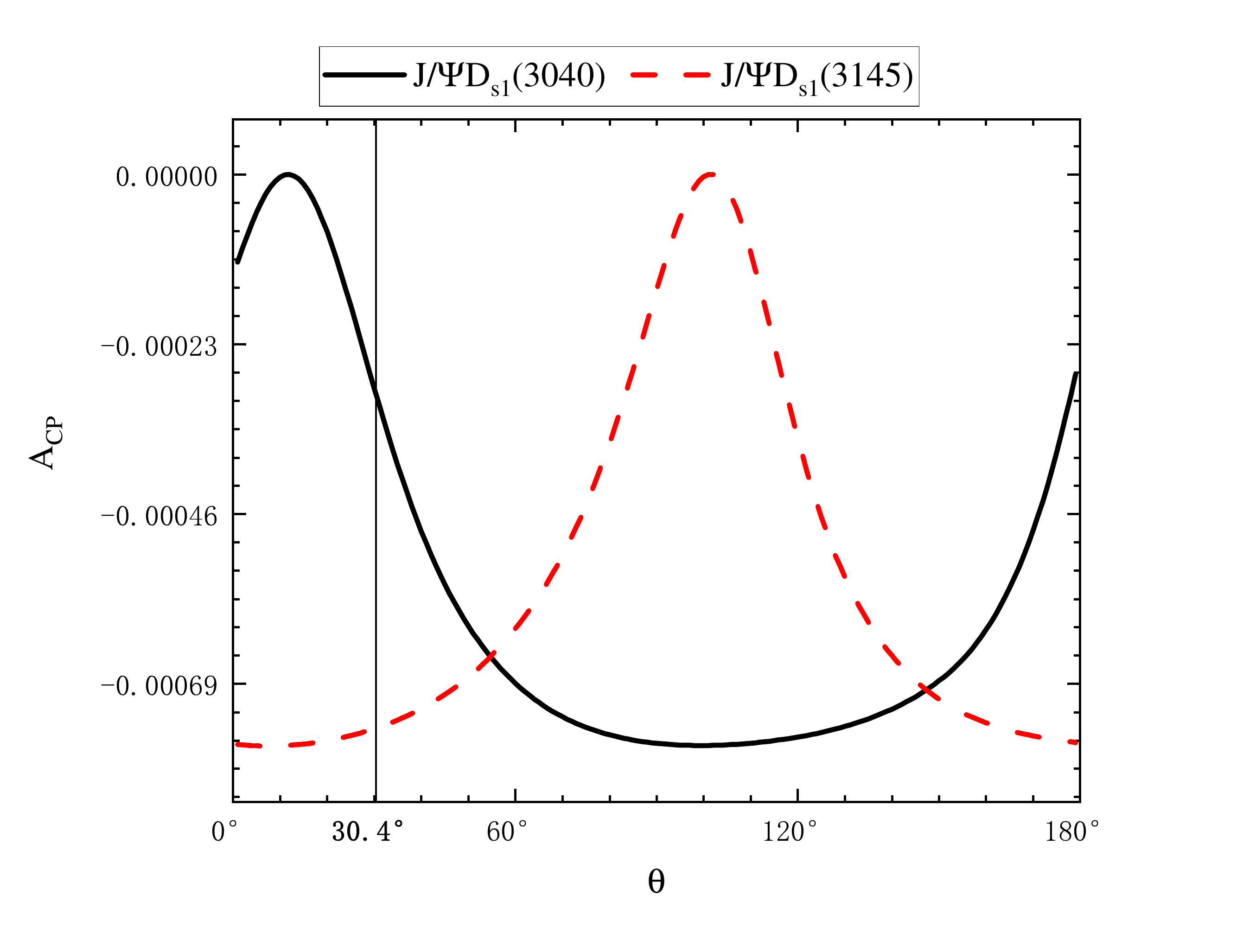}}\\
 \caption[]{$A_{cp}$ varies along with the mixing angle for the $B_c\to c\bar c(1S) + D_{(s)1}(2P)$.}
\end{figure}

\section{Acknowledgments}
This work was supported in part by the National Natural Science Foundation of China (NSFC) under Grant No.~12075073, 12047569, and 11865001. We also thank the HEPC Studio at Physics School of Harbin Institute of Technology for access to computing resources through INSPUR-HPC@hepc.hit.edu.cn.

\bibliographystyle{revtex}
\bibliography{bibleCP}

\end{document}